\documentclass[%
reprint,
 amsmath,amssymb,
 aps,
]{revtex4-2}

\usepackage{graphicx} % Include figure files
\usepackage{dcolumn}  % Align table columns on decimal point
\usepackage{bm}       % bold math
\usepackage{amssymb}  % additional math symbols
\usepackage{mathtools}
\usepackage[dvipsnames]{xcolor} % color for text
\usepackage[dvipsnames]{xcolor} % color for text
\usepackage{xcolor,colortbl}
\usepackage[capitalise,nameinlink]{cleveref} % multiple references
\usepackage{stfloats}
\usepackage{float}
\usepackage[normalem]{ulem}
\newcommand{\angstrom}{\textup{\AA}}
\newcolumntype{a}{>{\columncolor[gray]{0.75}}c}
\newcolumntype{b}{>{\columncolor[gray]{0.50}}c}
\newcolumntype{d}{>{\columncolor[gray]{0.80}}c}
\newcommand{\newtext}[1]{\textcolor{black}{#1}}

\begin{document}
\preprint{}

% % % % % % % % % % % % % % % % % % % % % % % % % % % % % % 
% Title
% % % % % % % % % % % % % % % % % % % % % % % % % % % % % % 
\title{Ultrafast control of material optical properties \textit{via} the infrared resonant Raman effect}% Force line breaks with \\
%\thanks{A footnote to the article title}%

% % % % % % % % % % % % % % % % % % % % % % % % % % % % % % 
% Authors and Affiliations
% % % % % % % % % % % % % % % % % % % % % % % % % % % % % % 
\author{Guru Khalsa}
 \email{guru.khalsa@cornell.edu}
 \affiliation{Department of Materials Science and Engineering, Cornell University, Ithaca, NY, 14853, USA}
 
\author{Nicole A. Benedek}
 \affiliation{Department of Materials Science and Engineering, Cornell University, Ithaca, NY, 14853, USA}

\author{Jeffrey Moses}
 \affiliation{School of Applied and Engineering Physics, Cornell University, Ithaca, NY, 14853, USA}

\date{\today}% It is always \today, today,
             %  but any date may be explicitly specified

% % % % % % % % % % % % % % % % % % % % % % % % % % % % % % 
% Abstract
% % % % % % % % % % % % % % % % % % % % % % % % % % % % % % 
\begin{abstract}
The Raman effect – inelastic scattering of light by lattice vibrations (phonons) – produces an optical response closely tied to a material’s crystal structure. Here we show that resonant optical excitation of IR and Raman phonons gives rise to a Raman scattering effect that can induce giant shifts to the refractive index and induce new optical constants that are forbidden in the equilibrium crystal structure. We complete the description of light-matter interactions mediated by coupled IR and Raman phonons in crystalline insulators -- currently the focus of numerous experiments aiming to dynamically control material properties -- by including a forgotten pathway through the nonlinear lattice polarizability. Our work expands the toolset for control and development of new optical technologies by revealing that the absorption of light within the terahertz gap can enable control of optical properties of materials over a broad frequency range.

\end{abstract}
\maketitle

% % % % % % % % % % % % % % % % % % % % % % % % % % % % % % 
% Introduction
% % % % % % % % % % % % % % % % % % % % % % % % % % % % % % 

The Raman effect arises from the inelastic scattering of light by phonons in crystals or vibrations in molecules. The simplest theories describing the effect assume that the optical polarizability -- that is, the general frequency-dependent polarizability of a crystalline material -- is predominantly electronic in origin\cite{boyd_chapter_2008,Wolverson1995}, and that it is a function only of the displacements of so-called Raman-active phonon modes. However, Herzberg \cite{gerhard_herzberg_chapter_1945} and Raman \cite{raman_vibration_1947} noted over 70 years ago that when infrared-active (IR) vibrational modes are driven to large amplitudes, nonlinear coupling between infrared- and Raman-active modes alters the optical polarizability. The relevant terms are complex and were difficult or impossible to verify and explore using the methods available at the time.\cite{barnes_structure_1935,geick_evidence_1965,keating_first-_1965,keating_higher_1965,davies_second-order_1968,maradudin_ionic_1970,wallis_ionic_1971,humphreys_ionic_1972,bendow_absorption_1973,sparks_infrared_1974}

The Raman effect has widespread scientific importance because it produces a basic optical response that is closely tied to a material's crystal structure. Raman spectroscopy exploits this tie and is an essential tool for understanding crystal symmetry, structure, and defects. For this reason, the effect predicted by Herzberg and Raman noted above -- changes to optical polarizability induced as a result of strongly driven IR vibrational modes
-- invites exploration as a potential tool for materials probing or control and, from an optics community perspective, also as a route for uncovering new ways to control optical properties through the interaction of laser light with material structure. The recent development of bright mid-infrared and THz sources capable of resonantly exciting collective modes in crystals has created new opportunities for exploring such novel optical phenomena \cite{kampfrath_coherent_2011,nicoletti_nonlinear_2016,hamm_perspective_2017}.

In this work, we show that resonant optical excitation of IR phonons strongly contributes to the optical polarizability via a Raman scattering mechanism mediated by the displacements of IR phonons, and that such excitations can be exploited to significantly modify a material's optical properties, including inducing new optical constants that are forbidden in the equilibrium crystal structure. This mechanism, which we refer to as infrared resonant Raman scattering  (IRRS), differs from the conventional Raman effect, in which typically only changes in the \emph{electronic} polarizability due to displacements of Raman phonons are considered. In contrast, IRRS involves the inelastic scattering of light and subsequent changes in the optical polarizability due to both changes in the electronic polarizability \emph{and} ionic polarizability due to Raman and IR phonon displacements. Additionally, IRRS is distinct from the ionic Raman scattering mechanism, which \emph{indirectly} produces an optical response through the coupling of phonons by the anharmonic lattice potential, rather than \emph{directly} through the nonlinear lattice polarizability. IRRS has a broad and distinct frequency response, which we find is orders of magnitude larger than ionic Raman scattering in an archetypal perovskite due to its direct nature.

Our work is significant in two general respects. First, selective optical pumping of IR phonons by frequency and field polarization direction can lead to control of the tensor components of the electric susceptibility, allowing control of effective linear and nonlinear optical responses of the material. This in turn can allow enhancement or suppression of fundamental optical properties such as absorption, birefringence, and electro-optic coefficient. Such control may aid the development of ultrafast optical technologies including optical switches, amplitude and phase modulators, and quantum optical logic gates. Second, we predict that the IRRS mechanism contributes to unidirectional displacement of Raman phonons and therefore changes of crystalline symmetry, giving rise to opportunities for ultrafast control of functional properties of crystalline materials tied to symmetry. This provides a pathway to ultrafast optical control of material symmetry and properties that is complementary to the so-called nonlinear phononics effect of phononic rectification, which has been the focus of much recent work for its ability to tune superconducting properties\cite{fausti_light-induced_2011,mankowsky_nonlinear_2014,mankowsky_optically_2017,cantaluppi_pressure_2018} and magnetic order\cite{wall_ultrafast_2009,forst_driving_2011,forst_spatially_2015,nova_effective_2017} in perovskites by intense laser pumping of an IR phonon.

To illustrate IRRS, we develop a semi-classical perturbation method approach in order to derive two-laser frequency contributions to the dielectric function from nonlinear contributions to the optical polarizability due to IR phonon displacements and IR-Raman phonon coupling. We further employ first-principles computational techniques (density functional theory) to investigate the effect in SrTiO$_3$, an archetypal insulating perovskite\cite{goodenough_electronic_2004} that has received recent attention in nonlinear phononics studies\cite{nova_metastable_2019,kozina_terahertz-driven_2019}. We find that the IRRS effect can be measured using a two-laser optical experiment where one laser is tuned to resonantly excite an IR active phonon. In addition to resonant enhancement of Stokes and anti-Stokes Raman peaks through IRRS, broad optical susceptibility changes are predicted that extend far above and below the IR phonon resonance. Furthermore, these changes are polarization direction-dependent and can result in modification of the number of optical axes and their directions. Additionally, we find that the mechanism responsible for IRRS induces unidirectional displacement of Raman active phonons. This presents an alternative pathway to the nonlinear phononics mechanism for quasistatic optical control of crystalline structure and properties.

In the following pages, we develop a theory of IRRS as follows. (1) We define a potential for describing a centrosymmetric crystalline lattice driven to large phonon mode displacements by IR light including the lowest-order nonlinear term in the lattice polarizability, which gives rise to IRRS. (2) We summarize the dependence of the linear electric susceptibiilty tensor on Raman phonon mode displacements and related symmetry considerations. The purpose of this section is to understand how basic optical properties and symmetries can be modified through control of specific Raman modes. (3) We employ a perturbation method approach to derive a third-order polarization and a Raman scattering susceptibility that capture the IRRS effect. These quantities may be used to explore the intensity-dependent modification of the linear susceptibilty tensor for a second laser as a result of resonant or near-resonant IR pumping by a first laser. (4) We employ first-principles computational techniques for the perovskite SrTiO$_3$ to investigate the physical mechanism of IRRS and opportunities for ultrafast material optical property control.

\section{Symmetry Considerations} \label{sec:Symmetry_Considerations}
%\label{section:Symmetry_Considerations}
\subsection{The lattice potential} \label{subsec:Lattice_Potential}

We start by expanding the lattice energy for the simpler case of a centrosymmetric crystal, in which IR and Raman modes have different symmetry (in noncentrosymmetric crystals, some phonons are both IR and Raman active; this lack of distinction between IR and Raman phonons leads to additional terms in the nonlinear polarizability and anharmonic potential, which we do not consider here). The lattice energy $U$ for the process we consider is then defined as,
	\begin{align}
		U_{lattice} &=  \frac{1}{2} K_{IR} Q_{IR}^2 + \frac{1}{2} K_{R} Q_{R}^2 \nonumber\\ 
		&- B Q_{IR}^2 Q_{R} - \Delta\vec{P}_{lattice}\cdot\vec{E}(t),
	   \label{eqn:potential_energy}
    \end{align}

\noindent
where $K_\sigma=M_{\sigma}\omega_\sigma^2 \ (\sigma = IR,R)$, is the effective spring constant of the IR and Raman phonons, respectively. $Q_{\sigma}$ is the real-space eigendisplacement of the phonon mode found by solving ${M \omega_\sigma^2 Q_\sigma= K Q_\sigma}$, where $M$ is the mass matrix, $K$ the force constant matrix at zero crystal momentum, and $\omega_\sigma$ is the corresponding phonon frequency. In addition to the harmonic energy associated with the IR and Raman phonons, we have included an anharmonic potential term that has been the focus of recent works on nonlinear phononics\cite{nicoletti_nonlinear_2016,mankowsky_non-equilibrium_2016,juraschek_thz_2019}. The last term contains the polarization change in the crystal, $\Delta\vec{P}$, due to optical excitation ($\vec{E}$).

There are both lattice and electronic contributions to $\Delta\vec{P}$. The lattice contribution to the polarization typically takes the form $\tilde{Z}^* Q_{IR}$, where $\tilde{Z}^*$ is the mode-effective charge induced by the displacement of the IR phonon\cite{ghosez_dynamical_1998}. Expanding $\Delta \vec{P}_{lattice}$ to second order in phonon amplitude for a centrosymmetric crystal we find,

\begin{align}
\Delta\vec{P}_{lattice} &=  \tilde{Z}^* Q_{IR} + b Q_R Q_{IR}.
\label{eqn:lattice_polarizability}
\end{align}

\begin{figure}[h]
\centering
\includegraphics[scale=0.5]{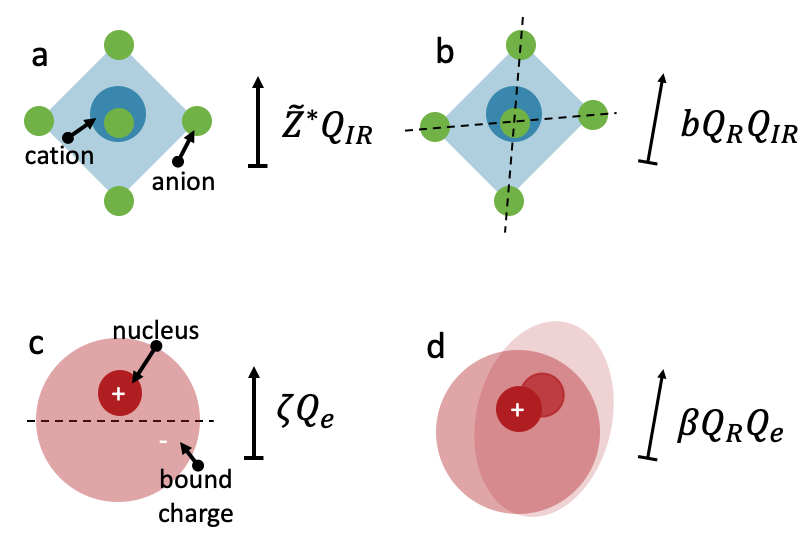}
\caption{\label{fig:nonlinear_dipole} Contributions to $\Delta\vec{P}$ in centrosymmetric crystals. (a) In an electric field, relative displacements of cations (blue) and anions (green) contribute to the lattice (ionic) polarizability. This is responsible for the dominant features of the low frequency dielectric response in the mid- and far- infrared. (b) Displacement of Raman phonons (shown, for example, as a distortion of the octahedral oxygen environment) alters the strength, and possibly direction, of the lattice polarizability. (c) A cloud of electrons bound to a nucleus displaces in response to an electric field. This is responsible for the high-frequency dielectric response. (d) Displacement of Raman phonons alters the electronic polarizability. This is the conventional Raman effect. A single atom is shown, although collective motion of Raman phonons is necessary for this effect.}
\end{figure}

\noindent
$Q_{IR}$ induces a dipole to which an electric field can be coupled directly. To preserve the dipole character of $\Delta \vec{P}$, $Q_{R}$ must transform as either a monopole or a quadrupole. Since the product of a dipole ($Q_{IR}$) and a monopole/quadrupole ($Q_R$) includes a dipole component, it follows that $bQ_{R}Q_{IR}$ is allowed in centrosymmetric crystals. Both terms of \cref{eqn:lattice_polarizability} are illustrated in \cref{fig:nonlinear_dipole}(a,b). The schematic shows that the first term, $\tilde{Z}^* Q_{IR}$, is produced by relative displacements of nuclei through the IR phonon.\cite{ghosez_dynamical_1998} Displacements of Raman phonons can change how the lattice polarizes through $Q_{IR}$ and this is captured through the nonlinear contribution, $b Q_R Q_{IR}$. The mathematical interpretation of this statement is that $b$ can be understood as the change in the mode-effective charge ($\tilde{Z}^*$) with respect to the Raman phonon. We note that the nonlinear lattice polarization $b Q_R Q_{IR}$ is in general a wave-vector dependent quantity so that ${b Q_{IR} Q_{R}\rightarrow b(\vec{q}_1 - \vec{q}_2) Q_1(\vec{q}_1)Q_2(-\vec{q}_2)}$ where we assume discrete translational symmetry in a crystal\cite{lax_infrared_1955}.
Since light cannot impart significant momentum on IR phonons, $\vec{q}_1 \approx \vec{0}$. It follows that $\vec{q}_2 = \vec{0}$. 

Previous experimental and theoretical works on the nonlinear phononics mechanism have focused on the anharmonic lattice potential ($BQ^2_{IR}Q_R$) and not nonlinear changes to the polarizability ($b Q_R Q_{IR}$).\cite{nicoletti_nonlinear_2016,mankowsky_non-equilibrium_2016,juraschek_thz_2019} In one recent exception, a nonlinear contribution to the polarizability $\Delta \vec{P}_{NL} \propto Q_{IR}^3$ was included in order to describe the optical changes in the Reststrahlen band of SiC when IR phonons were excited to large amplitude\cite{cartella_parametric_2018}. As we show below, including the lowest-order nonlinear contribution to the polarizability (\cref{eqn:lattice_polarizability}) is critical to understanding the optical response of a crystalline material in the mid-/far-infrared when IR phonons are strongly excited.

\subsection{The optical susceptibility} \label{subsec:Optical_Susceptibility}

To lowest order in a perturbative expansion of the polarizability in the electric field amplitude, the response of a crystalline material to light is captured by its frequency and wavevector-dependent linear susceptibility, $\chi^{(1)}(\omega,k)$, which connects the electric field of light to the induced polarization, $P_{\alpha} = \chi^{(1)}_{\alpha\beta} E_{\beta}$, where $\alpha$ and $\beta$ are Cartesian directions. 
For frequencies within the electronic gap of a centrosymmetric material, the susceptibility is dominated by the frequency-dependent electronic polarizability. $\chi^{(1)}_{\alpha\beta}(\omega)$ is a second-rank, polar tensor, which is a function of the amplitudes of particular lattice vibrations. 
In general, $\chi^{(1)}_{\alpha\beta}(\omega)$ can be decomposed into  spherically irreducible components (we use the notation of Ref. \onlinecite{hayami18}),
\begin{equation}
\chi^{(1)}_{\alpha\beta}(\omega) = \chi^{\text{(M)}} {\frac{1}{3}} \delta_{\alpha\beta} + \epsilon_{\alpha\beta\gamma} \chi^{\text{(D)}}_{\gamma} + \chi^{\text{(Q)}}_{\alpha\beta},
\label{eqn:susceptibility_irreducible}
\end{equation}
where $\delta$ is the Kronecker delta, $\epsilon$ is a Levi-Civita symbol, and $\chi^{\text{(M)}}$ are the scalar (monopole), $\chi^{\text{(D)}}_{\gamma}$ the vector (dipole), and $\chi^{\text{(Q)}}_{\alpha\beta}$ the deviator (quadrupole) contributions to the susceptibility:
\begin{align}
\chi^{\text{(M)}} &= \chi^{(1)}_{\alpha\alpha}(\omega) \\
\chi^{\text{(D)}}_{\gamma'} &= {\frac{1}{2}} \epsilon_{\alpha\beta\gamma'}  \chi^{(1)}_{\alpha\beta}(\omega) \\
\chi^{\text{(Q)}}_{\alpha\beta} &= {\frac{1}{2}}  (\chi^{(1)}_{\alpha\beta}(\omega) + \chi^{(1)}_{\beta\alpha}(\omega) ) - \chi^{\text{(M)}} {\frac{1}{3}} \delta_{\alpha\beta}.
\end{align}
Note, it is important to remember that while these irreducible components will always have the symmetry of the stated multipoles, the actual multipoles contributing to each can and usually are of higher order in the reduced (compared with free space) symmetry of a material's crystal field.\cite{paquet80} 

In this work, we are considering systems that have ground states that are centrosymmetric and time-reversal invariant (that is, non-magnetic and/or in zero-applied magnetic field). In such cases, Onsager's reciprocity relation dictates that $\chi^{(1)}_{\alpha \beta}(\omega) = \chi^{(1)}_{\beta \alpha}(\omega)$. It follows that the anti-symmetric tensor, $\chi^{\text{(D)}}_{\gamma'}$, vanishes by symmetry. In that case, the susceptibility is totally symmetric, 
$ \chi^{(1)}_{\alpha\beta}(\omega) = \chi^{\text{(M)}} {\frac{1}{3}} \delta_{\alpha\beta} + \chi^{\text{(Q)}}_{\alpha\beta}$, and the 1+5 independent components of $\chi^{(1)}_{\alpha\beta}(\omega)$ are written explicitly  as,
\begin{equation}
\left(
    \begin{matrix}
    ({\frac{1}{3}} \chi^{\text{(M)}} + \chi^{\text{(Q)}}_{xx}) &\chi^{\text{(Q)}}_{xy}& \chi^{\text{(Q)}}_{xz}\\
    \chi^{\text{(Q)}}_{xy}&   ({\frac{1}{3}} \chi^{\text{(M)}} + \chi^{\text{(Q)}}_{yy})  & \chi^{\text{(Q)}}_{yz}\\
  \chi^{\text{(Q)}}_{xz}&\chi^{\text{(Q)}}_{yz}& ({\frac{1}{3}} \chi^{\text{(M)}} + \chi^{\text{(Q)}}_{zz}) \\
    \end{matrix} \right).
    \end{equation}
The crystallographic symmetry of a particular material imposes additional constraints that dictate which elements are zero or non-zero, and how many components of $\chi^{(1)}_{\alpha\beta}(\omega)$ are independent. 

\subsection{The Raman tensor} \label{subsec:Raman_Tensor}

Now considering the Raman effect, we express the polarizability as a function of the amplitude, $Q_{i}$, of the lattice displacements associated with a particular phonon mode,  
\begin{equation}
    \label{polarizability_electronic}
    P_{\alpha} \approx \left(\chi^{(1)}_{0,\alpha\beta}(\omega) + Q_{i} \left. \frac{\partial \chi^{(1)}}{\partial Q_{i}} \right|_{Q_{i}\rightarrow 0} \right) E,
\end{equation}

where $\chi^{(1)}_{0,\alpha\beta}(\omega)$ is the susceptibility of the undistorted crystal and the second term describes the Raman effect. Note that $Q_{i}  \left. \left(\partial \chi^{(1)}/\partial Q_{i} \right)\right|_{Q_{i}\rightarrow 0}$ is constrained by the same symmetry as $\chi^{(1)}_{0,\alpha\beta}(\omega)$. This results in two distinct situations:\\ (1) $Q_i$ transforms as a monopole and therefore $\left. \left(\partial \chi^{(1)}/\partial Q_{i} \right)\right|_{Q_{i}\rightarrow 0}$ has the same non-zero elements as $\chi^{(1)}_{0,\alpha\beta}(\omega)$. In this case, the modulation of the nonlinear susceptibility can only affect the relative size of each term in $\chi^{(1)}_0$, but not induce new components.\\ (2) $Q_i$ transforms as a quadrupole and therefore  $\left. \left(\partial \chi^{(1)}/\partial Q_{i} \right)\right|_{Q_{i}\rightarrow 0}$ has new non-zero elements when compared to  $\chi^{(1)}_{0,\alpha\beta}(\omega)$. In this case, new optical constants are induced by displacing $Q_i$. 

In both of these cases, $Q_i$ is called a Raman-active phonon and hence we refer to the process described above as the infrared resonant Raman effect. In terms of \cref{eqn:susceptibility_irreducible}, displacing a monopolar Raman phonon alters nonzero elements of $\chi^{(M)}_{\alpha\beta}$ and $\chi^{(Q)}_{\alpha\beta}$ that already exist in the equilibrium structure. A quadrupolar Raman phonon lowers the symmetry of the crystal when displaced. As a result, the new induced optical constants -- constrained by symmetry to be zero at equilibrium -- are described by $\chi^{(Q)}_{\alpha\beta}$. 

We will use SrTiO$_3$ as a test material in this study. At temperatures below 105 K SrTiO$_3$ is a tetragonal crystal\cite{lytle_xray_1964} with space group symmetry $I4/mcm$ ($\#140$ - $D_{4h}$ point group). SrTiO$_3$ therefore has only two distinct nonzero optical constants, $\chi^{(1)}_{0,xx} = \chi^{(1)}_{0,yy} \ne\chi^{(1)}_{0,zz}$. The primitive unit cell contains 10 atoms and therefore has 30 phonons at the $\Gamma-$point that transform as, 

\begin{align*}
\Gamma&= \Gamma_1^+(A_{1g}) \\
&+\Gamma_2^+(B_{1g}) + 2 \Gamma_4^+(B_{2g})+3\Gamma_5^+(E_g) \\
&+4\Gamma_3^-(A_{2u})+6\Gamma_5^-(E_u)\\
&+2 \Gamma_3^+(A_{2g})\\ 
&+ \Gamma_1^-(A_{1u}) + \Gamma_2^-(B_{1u}), 
\end{align*}\noindent
where the first two lines describe the symmetries of the Raman-active phonons, the third line shows the symmetries of the IR-active phonons (which include acoustic modes), and the fourth and fifth lines are respectively even- and odd-symmetry silent phonons. The $\Gamma_1^+(A_{1g})$ phonon is the only monopolar Raman phonon. Displacing this phonon preserves the crystal symmetry and the number of optical constants. However, the relative size of each term in $\chi^{(1)}_0$ may change, as explained above. The remaining Raman-active phonons induce quadrupoles, and therefore new optical constants. As an example, consider the two  $\Gamma_4^+(B_{2g})$ Raman phonons, which transform like $xy$. Displacing either one of these quadrupolar phonons introduces new optical constants $\chi_{xy} = \chi_{yx}$, which induces, for example, a crystal polarization field along the $y-$direction when the crystal is illuminated with light polarized along the $x-$direction -- a property that is forbidden in the equilibrium structure. 

\section{Theoretical model} \label{sec:Theoretical_Model}
Having discussed symmetry principles that govern the infrared resonant Raman effect in crystals, we now present a theoretical model that can be used to describe it. While the infrared resonant Raman effect is expected to be present in small bandgap materials and metals, we focus our discussion and modeling on wide bandgap insulators in order to distill the physical implications of the infrared resonant Raman effect. This approach avoids obfuscation of the infrared resonant Raman effect from other considerations in metals (e.g. free charge screening) and semiconductors (e.g. Zener breakdown\cite{trompeter_visual_2006}, avalanche ionization\cite{pronko_avalanche_1998}).

\subsection{Equations of motion} \label{subsec:Equations_Of_Motion}

In order to derive the IRRS susceptibility we first derive the equations of motion from \cref{eqn:potential_energy,eqn:lattice_polarizability} in order to identify Raman scattering pathways. To capture the high-frequency response, we add the harmonic energy of an effective electronic coordinate, $Q_e$ (a phenomelogical parameter that describes the high-frequency dielectric response) and its coupling to the electric field and define a total potential $U = U_e + U_{lattice}$, where

\begin{equation}
	\label{eqn:electronic_energy}
    U_e = \frac{1}{2} K_e Q_e^2 - \Delta \vec{P}_e \cdot \vec{E}.
\end{equation}
\noindent
$K_e$ is an effective spring constant and the electronic polarizability includes coupling to the Raman phonon through,
\begin{equation}
	\label{eqn:electronic_polarizability}
    \Delta \vec{P}_e = \zeta Q_e + \beta Q_R Q_e.
\end{equation}

\noindent
Here $\zeta$ is an effective charge that describes the high-frequency optical response of the crystal. The coefficient $\beta$ describes the conventional electronically mediated Raman effect. \cref{fig:nonlinear_dipole}(c,d) shows schematic representations of linear and nonlinear electronic polarizability terms in \cref{eqn:electronic_polarizability}. Incidentally, comparing \cref{eqn:lattice_polarizability} and \cref{eqn:electronic_polarizability} we can see that since both $Q_{IR}$ and $Q_e$ transform as vectors, they both couple to Raman phonons in the same way. It follows that all of the symmetry arguments presented above are true for both the conventional electronically mediated Raman effect and the infrared resonant Raman effect.

Considering \cref{eqn:potential_energy,eqn:lattice_polarizability,eqn:electronic_energy,eqn:electronic_polarizability}, and taking derivatives of $U$ with respect to $Q_e$, $Q_{IR}$, and $Q_R$ we find the following equations of motion: 
\begin{subequations}
    \label{eqn:equations_of_motion_all}
    \begin{align}
        \mathcal{L}_{e} Q_{e}   &= \zeta E + \beta Q_{R} E\\
        \mathcal{L}_{IR} Q_{IR} &= \tilde{Z}^* E + b Q_R E + 2 B Q_{IR} Q_R\\
        \mathcal{L}_{R} Q_R     &= \beta Q_{e} E + b Q_{IR} E + B Q_{IR}^2.
    \end{align}
\end{subequations}
\noindent
We define the linear operator ${\mathcal{L}_{\sigma} = M_\sigma (\frac{d^2}{dt^2} + 2 \gamma_\sigma \frac{d}{dt} + \omega_{\sigma}^2)}$ for notational convenience. We have included a damping parameter $\gamma_\sigma$ and note that $M_\sigma$ is a reduced mass coordinate. The right-hand side of \cref{eqn:equations_of_motion_all} describes the driving terms through the electric field, coupling through the anharmonic lattice potential, and coupling through the nonlinear polarizability.

\begin{figure*}[!htbp]
\centering
\includegraphics[scale=0.55]{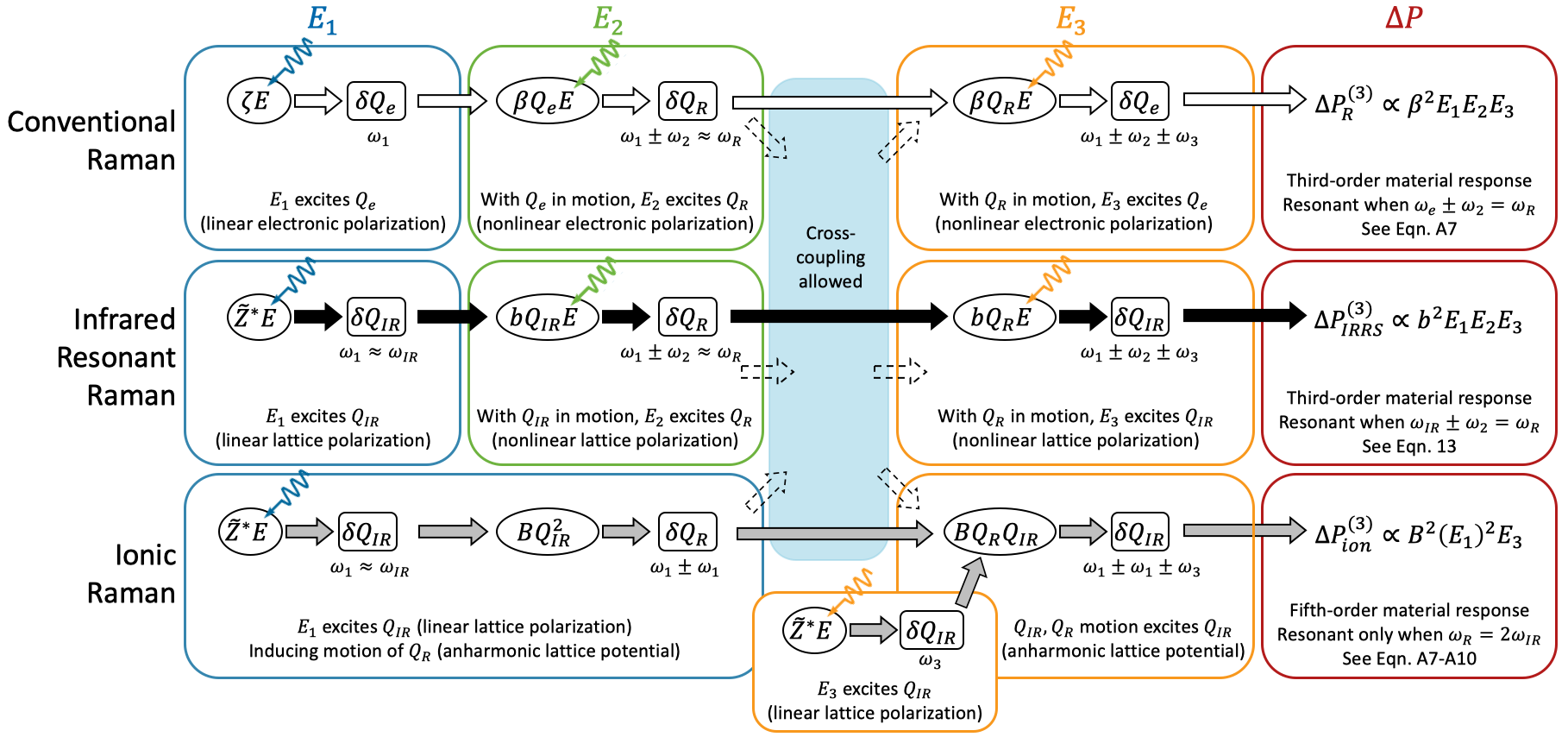}
\caption{
Comparison between the conventional (top), infrared resonant (middle), and ionic Raman scattering (bottom) mechanisms. Blue, green, and orange panels represent the effect of the three electric field components while the induced polarization is shown in the red panels. Driving terms in \cref{eqn:equations_of_motion_all} are shown in ovals with induced motion represented in rectangles with the allowed frequency response written below. The connection to linear and nonlinear polarization, and the anharmonic potential are shown in the text in each panel with resonance conditions for each effect called out in the polarization panels. Dashed arrows leading into the light-blue region in the middle of the figure show that cross-coupling is allowed between all three effects. The conventional and infrared-resonant Raman responses have potential for third-order resonance in the polarization (see \cref{eqn:chi_3_nonlinear_ionic_dipole,app_eqn:dipole_complete_solution,app_eqn:dipole_solution,app_eqn:third-order_susceptibility_general,app_eqn:third-order_susceptibility_two_fields}), while ionic Raman scattering has potential for a fifth-order resonance (see \cref{app_eqn:dipole_complete_solution,app_eqn:dipole_solution,app_eqn:third-order_susceptibility_general,app_eqn:third-order_susceptibility_two_fields}.)}
\label{fig:comparison}
\end{figure*}

Numerous scattering pathways present themselves through \cref{eqn:equations_of_motion_all}, as illustrated in \cref{fig:comparison}.

\textbf{Conventional (electronic) Raman scattering.} The conventional Raman scattering mechanism (top panel of \cref{fig:comparison}) is described by the $\beta$ coefficient. The effective electronic coordinate $Q_e$ can respond to any frequency of light at or below its fundamental frequency $\omega_e$. The combined motion of $Q_e$ with the electric field induces a driving force for the Raman coordinate $Q_R$ through the parameter $\beta$ (first term on the right-hand side of \cref{eqn:equations_of_motion_all}c).

\textbf{Infrared resonant Raman scattering.} The IRRS mechanism (middle panel of \cref{fig:comparison}) is analogous to the conventional Raman scattering pathway, except it substitutes the displacement of the electron cloud, characterized by $Q_e$, with the displacement of an IR phonon, $Q_{IR}$. Displacement of an IR phonon in the presence of an electric field induces displacements of Raman phonons through the nonlinear polarizability, described by the coefficient $b$. The largest IR phonon response is when a component of the applied field is tuned to the frequency $\omega_{IR}$. Once resonantly excited, the combined motion of $Q_{IR}$ and the electric field drive $Q_R$ through $b$. 

\textbf{Ionic Raman scattering.} An additional Raman scattering mechanism (bottom panel of \cref{fig:comparison}) is provided by the anharmonic lattice potential. After resonant excitation of $Q_{IR}$ there is a driving force on $Q_R$ induced by the parameter $B$ proportional to $Q_{IR}^2$. This scattering pathway is an example of nonlinear ionic source terms in the dielectric response arising from the anharmonic lattic potential,\cite{Armstrong:62} and has been described in the literature as ionic Raman scattering.\cite{maradudin_ionic_1970,wallis_ionic_1971,humphreys_ionic_1972,juraschek_sum-frequency_2018} It should be noted that this pathway is distinct from the other two in that the applied field only directly induces a displacement to an IR phonon, i.e., the presence of an electric field is not required to couple IR phonon motion to Raman phonon motion. It should also be noted that this prevents resonant excitation of the Raman phonon except for the singular case where the frequency of the applied electric field is half the frequency of the Raman phonon. In contrast, in conventional Raman scattering and in IRRS, resonant Raman excitation may be achieved through the combination of any two laser frequencies such that $\omega_1 \pm \omega_2 \approx \omega_R$.

\subsection{Perturbative approach to the nonlinear susceptibility} \label{sec:Perturbation_Theory}

We derive expressions for the optical susceptibility using a perturbative approach in the parameters $\beta$, $b$, and $B$ and solve \cref{eqn:equations_of_motion_all} in the presence of a multicomponent electric field, expressing $Q_\sigma$ in terms of components of $E$. These solutions are used to remove $Q_e$, $Q_{IR}$, and $Q_R$ from \cref{eqn:lattice_polarizability,eqn:electronic_polarizability} and express the polarizability in terms of only the electric field so that the linear and nonlinear contributions to the polarizability can be collected. By symmetry, we can see that the dipole active coordinates $Q_e$ and $Q_{IR}$ will be replaced by a parallel electric field $E$, while each Raman phonon will be replaced by two factors of the electric field. From \cref{eqn:lattice_polarizability,eqn:electronic_polarizability} we see that $\tilde{Z}^* Q_{IR}$ and $\zeta Q_e$ will contribute to first-order in $E$, while $b Q_{R} Q_{IR}$ and $\beta Q_R Q_e$ will contribute terms to third-order in $E$. In this way, we find that the polarizability includes both linear and third-order susceptibilities $\vec{P} = \chi^{(1)}\vec{E} + \chi^{(3)} \vec{E} \vec{E} \vec{E}$.  

To simplify the presentation we focus on a scenario where one laser is used to drive an IR phonon on, or near, resonance while a second laser samples the optical response at another frequency taken to vary over a large range. This can be viewed as intensity dependent dielectric changes due to the excitation of an IR phonon. The electric field is taken to be
\begin{equation}
    \label{eqn:two_fields}
    E (t)= \frac{1}{2} \left( E_1 e^{-i \omega_1 t} + E_{-1} e^{i \omega_1 t} + E_2 e^{-i \omega_2 t} +  E_{-2} e^{i \omega_2 t} \right)
\end{equation}
\noindent
with the field strength defined with the constraint $E_{-n} = E_n^*$ to enforce real fields. Notice that with this convention we can take $\omega_{-n} = -\omega_n$ and define the general multi-component electric field as $E(t)=\frac{1}{2}\sum_{n=\pm1,\pm2,...}E_n e^{-i\omega_n t} $ by summing over both negative and positive components of the field. The third-order susceptibility for the process under investigation, $\chi^{(3)}(\omega_2;\omega_1,-\omega_1,\omega_2) \equiv \chi^{(3)}(2;1,-1,2)$, describes the macroscopic nonlinear polarizability at $\omega_2$ due to the mixing of light at $\omega_1$ (taken to be near the IR resonance) and $\omega_2$. To simplify $\chi^{(3)}$ further we assume that all laser frequencies are far below the electronic resonance $\omega_e$, but include general expressions for the polarizability in the Appendix (see \cref{app_eqn:dipole_complete_solution}). The nonlinear polarization of the crystal is then defined by $P^{(3)} =\chi^{(3)}(2;1,-1,2) |E_1|^2 E_2 e^{-i\omega_2 t} + c.c. $ $\equiv \delta \chi^{(1)}(I_1) E_2 e^{-i\omega_2 t} + c.c.$ We see that the polarization is measured at the frequency $\omega_2$ and can be interpreted as an intensity dependent change to the \emph{linear} susceptibility $\delta \chi^{(1)}(\omega_2;I_1) \propto I_1$, where the intensity $I_1=|E_1|^2$. With this definition, the effective linear susceptibility at $\omega_2$ takes the form $\chi^{(1)}_{eff}(\omega_2;I_1) = \chi^{(1)}(\omega_2)+\delta \chi^{(1)}(\omega_2;I_1)$, which reflects that the linear optical properties for $E_2$ are altered by the presence of a strong $E_1$.

Even with these assumptions the perturbative approach leaves us with cumbersome expressions. Focusing on the component of the third-order susceptibility due solely to the nonlinear contribution to the polarizability with coefficient $b$ (corresponding to the IRRS effect introduced here, see middle pathway in \cref{fig:comparison}), we define and find
\begin{widetext}
\begin{equation}
    \label{eqn:chi_3_nonlinear_ionic_dipole}
    \begin{split}
        \chi^{(3)}_{b,b}(2;1,-1,2) 
            = \frac{(b \tilde{Z})^2}{\epsilon_0 V} &\times \left[ 2 \left(G^{IR}_1 + G^{IR}_{-1} + G^{IR}_2 + G^{IR}_{-2} \right) G^{IR}_2 G^R(\omega_{1}-\omega_{1}=0) \right. \\
            &+ \left( G^{IR}_2 G^{IR}_2 + G^{IR}_1 G^{IR}_2 + G^{IR}_2 G^{IR}_{-1} + G^{IR}_1  G^{IR}_{-1} \right) G^{R}(-\omega_1 + \omega_2) \\
            &\left.+ \left( G^{IR}_2 G^{IR}_2 + G^{IR}_1 G^{IR}_2 + G^{IR}_2 G^{IR}_{-1} + G^{IR}_1  G^{IR}_{-1} \right) G^{R}(\omega_1 + \omega_2) \right].\\
%        \chi^{(3)}_{B,B}(2;1,-1,2) 
  %          &= \frac{\tilde{Z}^4 B^2}{\epsilon_0 V} \times \\    
    %        &\left\{ 2 \left(G^{IR}_1 G^{IR}_{-1} + G^{IR}_{2} G^{IR}_{-2}\right)(G^{IR}_2)^2 G^R(\omega_{mn}=0) \right. \\
%            &+ 2 \left(G^{IR}_1 G^{IR}_{-1}\right)(G^{IR}_2)^2 G^R(-\omega_1+\omega_2)  \\
  %          &+ 2 \left(G^{IR}_1 G^{IR}_{-1}\right)(G^{IR}_2)^2 G^R(\omega_1+\omega_2)  \\
    \end{split}
\end{equation}
\end{widetext}
\noindent
In this expression $G_n^{\sigma} = G^{\sigma}(\omega_n)$ is the inverse of the linear-operator $\mathcal{L}_{\sigma}$ in the frequency domain (see \cref{eqn:equations_of_motion_all}), evaluated at $\omega_n$. That is,
%
%\begin{equation}
%   \label{eqn:green_func}
%   G^\sigma_n = G^\sigma (\omega_n) = \frac{1/M_\sigma}{\left( %\omega_{\sigma}^2 - \omega_n^2 \right) - 2 i \gamma_{\sigma} \omega_n }.
%\end{equation}
\begin{subequations}
   \begin{align}
   \label{eqn:green_func}
   G^{IR}_n = G^{IR} (\omega_n) &= \frac{1/M_{IR}}{\left( \omega_{IR}^2 - \omega_n^2 \right) - 2 i \gamma_{IR} \omega_n } \\           
   G^R_{mn} = G^R (\omega_{mn}) &= \frac{1/M_R}{\left( \omega_{R}^2 - \omega_{mn}^2 \right) - 2 i \gamma_{R} \omega_{mn} }.
    \end{align}
\end{subequations}
\noindent
where $\omega_{mn} = \omega_m + \omega_n$. \cref{eqn:chi_3_nonlinear_ionic_dipole} is organized by the frequency response of the Raman phonon. The right-hand side of the first line shows the contribution from \emph{static} displacement of the Raman phonon through rectification of the electric field components ($\omega_{mn} = 0$, remembering that $\omega_{n(m)}$ is summed over positive and negative frequencies). The remaining lines are due to dynamical responses of the Raman phonon. The second line gives the contribution from the motion of the Raman phonon at $-\omega_1 + \omega_2$. Since the function $G^R$ is peaked at the Raman frequency, the second line of \cref{eqn:chi_3_nonlinear_ionic_dipole} is resonant when $(-\omega_1+\omega_2) = \pm \omega_R$, while the third line is resonant when $(\omega_1+\omega_2) = \pm \omega_R$.

Each term in \cref{eqn:chi_3_nonlinear_ionic_dipole} includes two contributions from $G^{IR}$ and one from $G^R$. This corresponds to a double-resonance due to the IR phonon response and a single resonance from the Raman response. An overall \textit{triply resonant} response is therefore possible in the two-laser scenario we are describing. As an example, consider a first laser resonant with the IR phonon ($\omega_1 = \omega_{IR}$) and a second laser tuned to a frequency $\omega_2$ away from the IR resonance so that $-\omega_1 + \omega_2 = \pm\omega_R$, as is expected with a Stokes($-\omega_R$)/anti-Stokes Raman($+\omega_R$) response. We note that this triply resonant response is analogous to conventional resonant stimulated Raman scattering, wherein the first laser is tuned to an electronic resonance ($\omega_1=\omega_e$). However, these responses are found in disparate ranges of the optical spectrum.

\cref{fig:chi3_schematic} shows the magnitude and phase of $\chi^{(3)}(2;1,-1,2)$ schematically when coupling between a single IR phonon at $f_{IR} = \omega_{IR}/2\pi = 20\ \text{THz}$ and a single Raman phonon at $f_{R} = \omega_{R}/2\pi  = 12\ \text{THz}$ is allowed through the nonlinear dipole moment. The curves indicate the contribution from $\chi^{(3)}_{b,b}$, \textit{i.e.}, from IRRS. Two peaks are seen in the magnitude of $\chi^{(3)}_{b,b}$ at the difference- and sum-frequencies of $8\ \text{THz}$ and $32\ \text{THz}$, respectively. While the magnitude remains constant below $8\ \text{THz}$, it falls off as $f^{-2}$ for high-frequency (above $32\ \text{THz}$). At the difference-frequency the phase of $\chi^{(3)}$ approaches $-\pi/2$, suggesting $\chi^{(3)}_{b,b}$ is purely imaginary and that there will be emission of radiation at $8\ \text{THz}$. At the sum-frequency the phase approaches $+\pi/2$, suggesting absorption of radiation at $32\ \text{THz}$. This is consistent with the classical theory of Stokes and anti-Stokes Raman scattering. We therefore find that the nonlinear contribution to the polarization shown in \cref{eqn:lattice_polarizability} resonantly enhances the Stokes and anti-Stokes response known from the conventional Raman effect through excitation of an IR phonon. 

In between the difference- and sum-frequencies the phase of the response is zero, indicating that the real part of the dielectric constant will increase through $\chi^{(3)}_{b,b}$. Below the difference-frequency, and above the sum-frequency, the phase of $\chi^{(3)}_{b,b}$ becomes $\mp\pi$ suggesting a decrease in the real part of the dielectric constant in these frequency ranges. These effects are also consistent with conventional Raman scattering.

\section{First-principles calculations} \label{sec:First_Principles}
The model described in the previous section requires various inputs, which we obtain from first-principles calculations. Density functional theory calculations were performed using projector augmented wave potentials and the local density approximation (LDA), as implemented in VASP. We use a 600 eV plane wave cut-off and a $6\times 6\times 4$ Monkhorst-Pack $\mathbf{k}$-point grid for a $\sqrt{2}a\times\sqrt{2}a\times 2a$ supercell, where $a$ is the optimized lattice constant of SrTiO$_3$ in the cubic $Pm\bar{3}m$ phase. The calculated lattice constants $a = b = 5.44$ \AA \ and $c = 7.75$ \AA\ underestimate the experimental\cite{kiat_rietveld_1996} lattice constants  of $a = b = 5.51$ \AA \ and $c = 7.81$ \AA\ as expected in LDA. Phonons and Born effective charges were calculated using density functional perturbation theory within VASP and used to calculate dynamical mode effective charges. Further details are given in Ref. \onlinecite{khalsa18}.

For the anharmonic potential terms $B$ we compute the total energy $U_{lattice}$ on a mesh of IR and Raman phonon amplitudes up to $\pm 20$ pm in $2$ pm steps and fit to a symmetry constrained polynomial. $B$ is then interpreted as the derivative of the total energy fit with respect to the real-space eigendisplacements of each phonon $B=\frac{\partial^3 U}{\partial Q_{IR}^2 \partial Q_{R}}$. Similarly, the coefficient controlling the strength of the nonlinear contribution to the lattice polarization, $b$, is evaluated by taking derivatives of a symmetry constrained polynomial fit to the polarization $b=\frac{\partial \Delta \vec{P}}{\partial Q_{IR} \partial Q_{R}}$ calculated using the modern theory of polarization.\cite{Resta2007}

The damping parameters $\gamma_\sigma$ are temperature dependent quantities. Because they require evaluation of, minimally, all third-order force constants, and often all electron-phonon coupling pathways and defect scattering pathways, their calculation from first-principles is prohibitively expensive. Instead of calculating damping terms from first principles, we have explored a range of values, 0.1 THz -- 1.0 THz, consistent with typical damping parameters for phonons. In what follows, we fix damping parameters to 0.1 THz, corresponding to 10 ps lifetimes, and comment on any strong dependence of the optical response on this value when necessary.

\section{Results} \label{sec:Results}
%\subsection{Third-order suceptibility in SrTiO$_3$} \label{subsec:Third_Order_Susceptibility}

\begin{figure}[h!]
\centering
\includegraphics[scale=0.3]{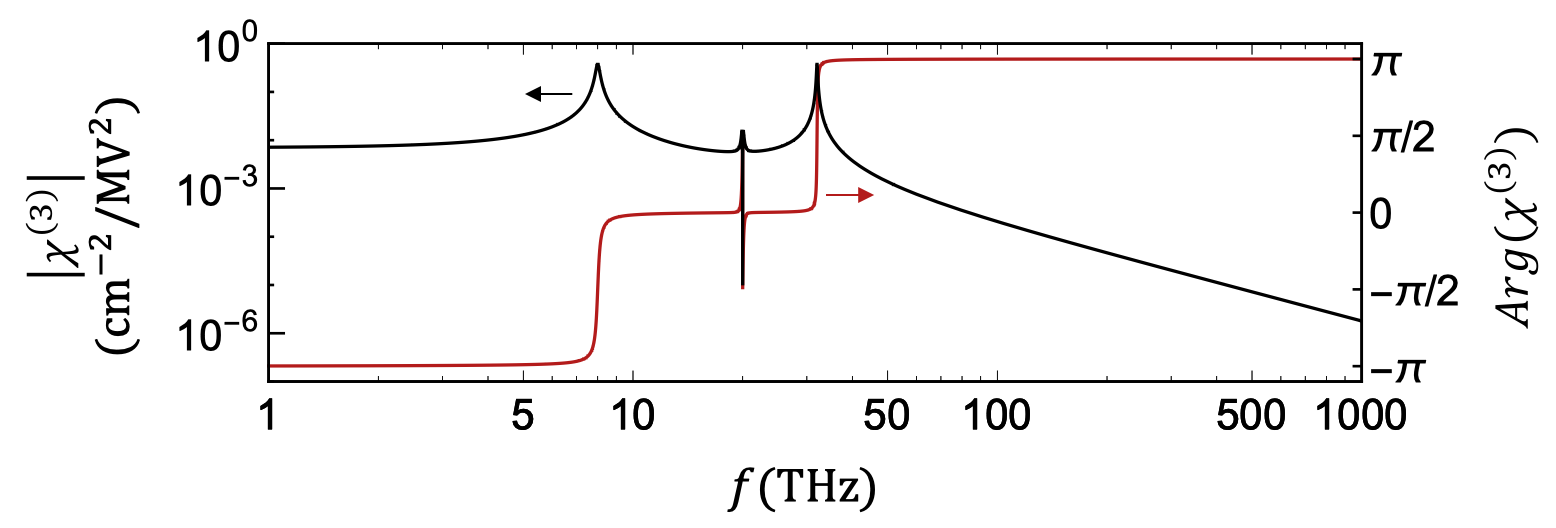}
\caption{\label{fig:chi3_schematic} Magnitude (black) of the third-order IRRS susceptibility, $\chi^{(3)}_{b,b}$, resulting from $bQ_{IR}Q_R$ for an IR phonon at $20\ \text{THz}$ with effective charge $\tilde{Z}^*=1\ e$ coupled to a Raman phonon at $12\ \text{THz}$ with a coupling strength $b= 1\ e/\angstrom$. The sum- and difference-frequencies are $32\ \text{THz}$ and $8\ \text{THz}$, respectively. The phase (red) is referenced to the right vertical axis. %The amplitude of the third-order susceptibility induced through the anharmonic lattice potential only ($BQ^2_{IR}Q_R$, dashed blue), with $B=10$ eV/$\angstrom^3$, is shown to be many orders of magnitude smaller than the infrared resonant Raman response.
}
\end{figure}

We employed the theoretical model of Section \ref{sec:Theoretical_Model} together with the first-principles calculations of Section \ref{sec:First_Principles} to investigate the changes to optical material properties that can be induced in SrTiO$_3$ through IRRS. As we are most interested in the changes to optical constants and optical symmetry that arise from resonant or near-resonant IR pumping, we explored cubic susceptibility terms, in the form of the intensity-dependent change to the linear susceptibility, for an applied field $\vec{E}_2$ due to the application of an infrared-resonant field $\vec{E}_1$, $\delta \chi^{(1)}_{ij}(\omega_2;I_1(\omega_1)) = \chi^{(3)}_{ijji}(2;1,-1,2)|E_{1,j}|^2$, as defined in Section \ref{sec:Perturbation_Theory}. The symmetry of $\chi^{(3)}$ for the infrared resonant Raman response depends on the symmetry of the displaced Raman phonons as discussed in Section \ref{sec:Symmetry_Considerations}. The frequency dependence of the tensor components of $\chi^{(3)}$, as will be seen below, has the expected form of a Raman scattering susceptibility as seen in \cref{fig:chi3_schematic}, with Stokes and anti-Stokes peaks corresponding to differences and sums of the IR and Raman phonons of the material, respectively. The response, however, is complicated by the presence of multiple IR-Raman phonon couplings for each activated IR phonon and the different $\chi^{(3)}$ symmetries corresponding to each Raman phonon. As a result, the intensity dependent dielectric response through the excitation of the IR phonons (see \cref{eqn:chi_3_nonlinear_ionic_dipole} and discussion) and the resulting changes to the optical symmetry are intimately tied to the frequency range spanned in the experiment and the polarizations of the applied fields.

\begin{figure}[h]
\centering
\includegraphics[scale=0.32]{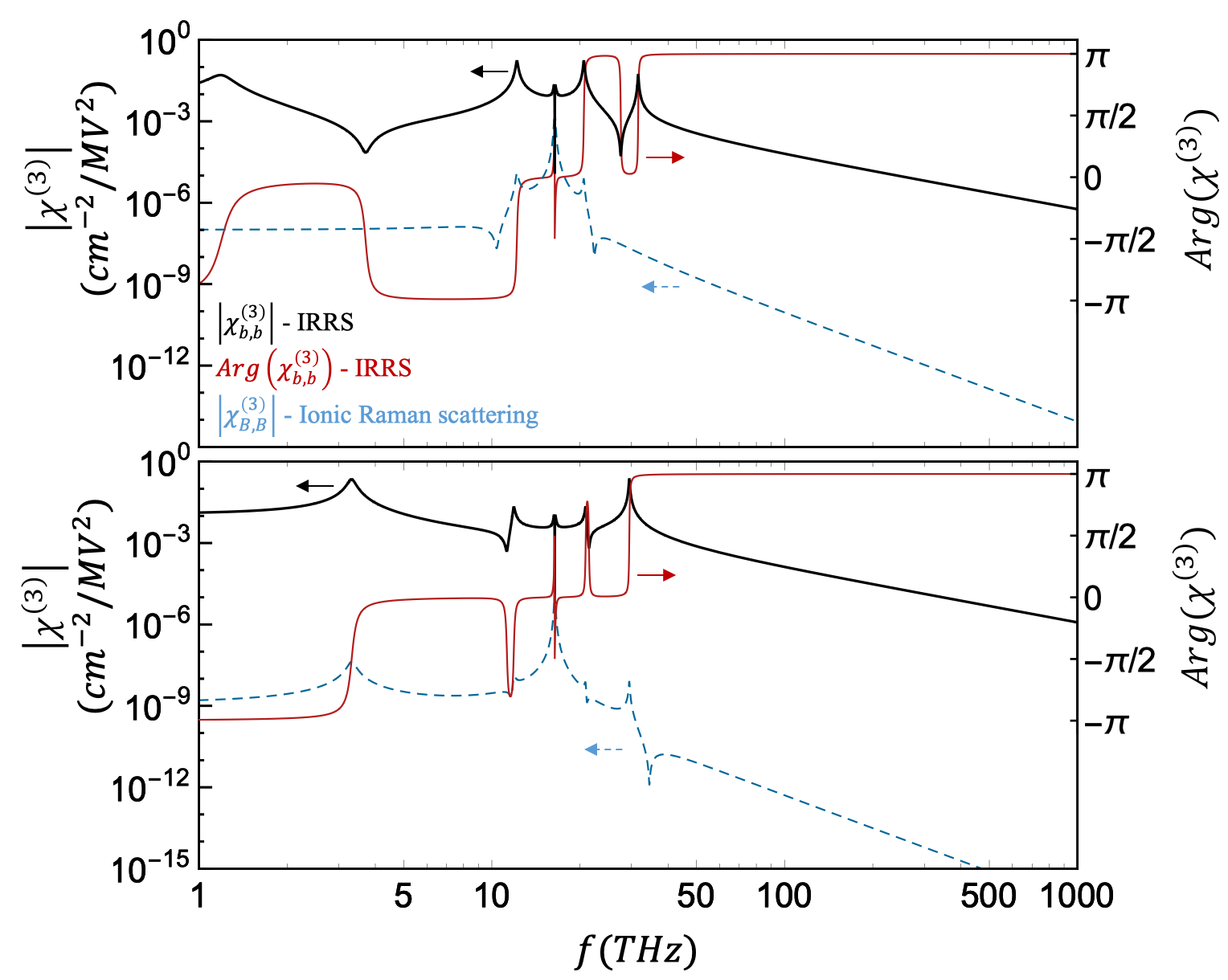}
\caption{Simulated magnitude (black) and phase (red) of the third-order susceptibility due to IRRS, $\chi^{(3)}_{b,b}$, in SrTiO$_3$ when the highest frequency $\Gamma_5^-(E_u)$ IR phonon is resonantly excited (${f_{IR}=16.40\ \text{THz}}$). (a) $\chi^{(3)}_{xxxx}(2;1,-1,2)$ exhibits coupling to the $\Gamma_1^+(A_{1g})$ and $\Gamma_2^+(B_{1g})$ Raman phonons. The difference- and sum-frequencies that identify peaks are $12.18\ \text{THz}$ and $20.62\ \text{THz}$ for the $\Gamma_1^+(A_{1g})$ phonon, and $1.19\ \text{THz}$ and $31.61\ \text{THz}$ for the $\Gamma_2^+(B_{1g})$ phonon. (b) $\chi^{(3)}_{yxxy}(2;1,-1,2)$ exhibits coupling to the two $\Gamma_4^+(B_{2g})$ phonons. The difference-frequencies are $3.38\ \text{THz}$ and $11.88\ \text{THz}$, while the sum-frequencies are $29.42\ \text{THz}$ and $20.92\ \text{THz}$. Also shown (dashed), the corresponding magnitude of the third-order susceptibility due to the anharmonic lattice potential (ionic Raman scattering), $\chi^{(3)}_{B,B}$.} 
\label{fig:chi3_GM5-_high}
\end{figure}

\textbf{Optical symmetry breaking.} \cref{fig:chi3_GM5-_high} shows the simulated third-order susceptibility $\chi^{(3)}(2;1,-1,2)$ for SrTiO$_3$ when $\vec{E}_1$ is polarized along the $x-$direction (taken to be parallel to the crystallographic $a-$axis) and resonant with the $\Gamma_5^-(E_u)$ IR phonon at $f_{IR} = 16.40 \text{THz}$. The nonlinear susceptibility is notably more complex than that of the simplified response in \cref{fig:chi3_schematic}. This is due to the coupling to multiple Raman-active phonons of SrTiO$_3$. Through these IR-Raman couplings mediated by IRRS, intensity-dependent changes to the effective linear susceptibility for a second field $\vec{E}_2$ occur for $\vec{E}_2$ fields polarized along $x$, $\delta \chi^{(1)}_{xx}(\omega_2;I_1) = \chi^{(3)}_{xxxx}(2;1,-1,2)|E_{1,x}|^2$, and along $y$, $\delta \chi^{(1)}_{yy}(\omega_2;I_1) = \chi^{(3)}_{yxxy}(2;1,-1,2)|E_{1,x}|^2$.

We first consider the case where both $\vec{E}_1$ and $\vec{E}_2$ are polarized along the $x-$axis. The Raman phonons that contribute to the optical response in this case (via $\chi^{(3)}_{xxxx}(2;1,-1,2)$) are the $\Gamma_1^+(A_{1g})$ mode -- which is fully symmetric -- and the $\Gamma_2^+(B_{1g})$ which transforms as $x^2-y^2$ (\cref{fig:chi3_GM5-_high}a). The resonance peaks are then the sums (anti-Stokes peaks) and differences (Stokes peaks) of the IR resonance and the $\Gamma_1^+(A_{1g})$ and $\Gamma_2^+(B_{1g})$ Raman phonons. We see peaks at $12.18\ \text{THz}$ and $20.62\ \text{THz}$ through coupling to $\Gamma_1^+(A_{1g})$, and $1.19\ \text{THz}$ and $31.61\ \text{THz}$ through coupling to $\Gamma_2^+(B_{1g})$ Raman phonons. The phase of the nonlinear suceptibility is also complicated, showing multiple absorption and emission pathways. However, there are broad spectral regions where  the IRRS susceptibility remains purely real (that is, when $Arg(\chi^{(3)})=\{0,\pm \pi\}$) in ranges of frequencies not near resonances.

The intensity-dependent changes to the effective linear susceptibilty induced by the $\Gamma_1^+(A_{1g})$ Raman phonon preserve the uniaxial optical symmetry. That is, $\delta\chi^{(1)}_{xx}(\omega_2,I_1) = \delta\chi^{(1)}_{yy}(\omega_2,I_1) \ne \delta\chi^{(1)}_{zz}(\omega_2,I_1)$, a result of $\chi^{(3)}_{xxxx}(2;1,-1,2)|E_{1,x}|^2 = \chi^{(3)}_{yyyy}(2;1,-1,2)|E_{1,y}|^2 \ne \chi^{(3)}_{zzzz}(2;1,-1,2)|E_{1,z}|^2$. However, the response of the $\Gamma_2^+(B_{1g})$ Raman phonon alters the linear susceptibility so that $\delta\chi^{(1)}_{xx}(\omega_2,I_1)\ne \delta\chi^{(1)}_{yy}(\omega_2,I_1)$ -- which is the property of a biaxial optical symmetry -- since $\chi^{(3)}_{xxxx}(2;1,-1,2)|E_{1,x}|^2 \ne \chi^{(3)}_{yyyy}(2;1,-1,2)|E_{1,y}|^2$. Since both Raman modes are driven by the IR mode simultaneously through IRRS and thus their effects on the optical response cannot be disentangled, the net result of resonant pumping of the $\Gamma_5^-(E_u)$ IR phonon is a shift from uniaxial to biaxial symmetry, with the angle that splits the two optical axes controlled by $I_1$.

If we instead consider the case where $\vec{E}_2$ is polarized along the $y-$axis (and the $\Gamma_5^-(E_u)$ IR phonon at $f_{IR} = 16.40 \text{THz}$ is still pumped by a $\vec{E}_1$ polarized along the $x-$axis), we now observe a cross-polarization effect. The two $\Gamma_4^+(B_{2g})$ Raman phonons at $13.02\ \text{THz}$ and $4.52\ \text{THz}$ now facilitate the response through $\chi^{(3)}_{yxxy}(2;1,-1,2)$ (\cref{fig:chi3_GM5-_high}b). Again, complicated amplitude peaks and phase responses are seen with difference-frequencies (Stokes peaks) and sum-frequencies (anti-Stokes peaks) of $3.38\ \text{THz}$ and $11.88\ \text{THz}$, and $29.42\ \text{THz}$ and $20.92\ \text{THz}$, respectively. Interpreting this response from the perspective of the intensity dependent linear susceptibility, we find that $\delta\chi^{(1)}_{xy}(\omega_2;I_1) = \delta\chi^{(1)}_{yx}(\omega_2;I_1)$ are nonzero. That is, an optical constant not allowed by the symmetry of the equilibrium structure appears and becomes intensity dependent. This implies an intensity dependent shift to the principal optical axes.

 \Cref{fig:chi3_GM5-_high} also shows that the contribution to $\chi^{(3)}(2;1,-1,2)$ coming purely from the anharmonic lattice potential (blue, dashed line in \cref{fig:chi3_GM5-_high}) is many orders of magnitude smaller than the contribution due to IRRS in SrTiO$_3$. This could have been anticipated by the definition of the induced polarization (\cref{eqn:electronic_polarizability,eqn:lattice_polarizability}) and the equations of motion {\cref{eqn:equations_of_motion_all}} along with the perturbation approach shown schematically in the bottom-panel of \cref{fig:comparison} and developed in Sec. \ref{sec:Perturbation_Theory}. In the absence of $b$ and $\beta$ the only possible changes to the polarization are through changes in $Q_{IR}$. While the anharmonic lattice potential induces a first-order change in the Raman coordinate $Q_R$, it only induces a second-order change in $Q_{IR}$. Thus, we should expect the pure anharmonic lattice contribution to the third-order susceptibility to be small. An expression for the changes to $\chi^{(3)}$ induced by the anharmonic lattice potential and all possible cross-coupling pathways for $\chi^{(3)}$ up to second-order in perturbation theory are given in the Appendix (see \cref{app_eqn:third-order_susceptibility_general,app_eqn:third-order_susceptibility_two_fields}).

\begin{figure*}[h!]
\centering
\includegraphics[scale=0.43]{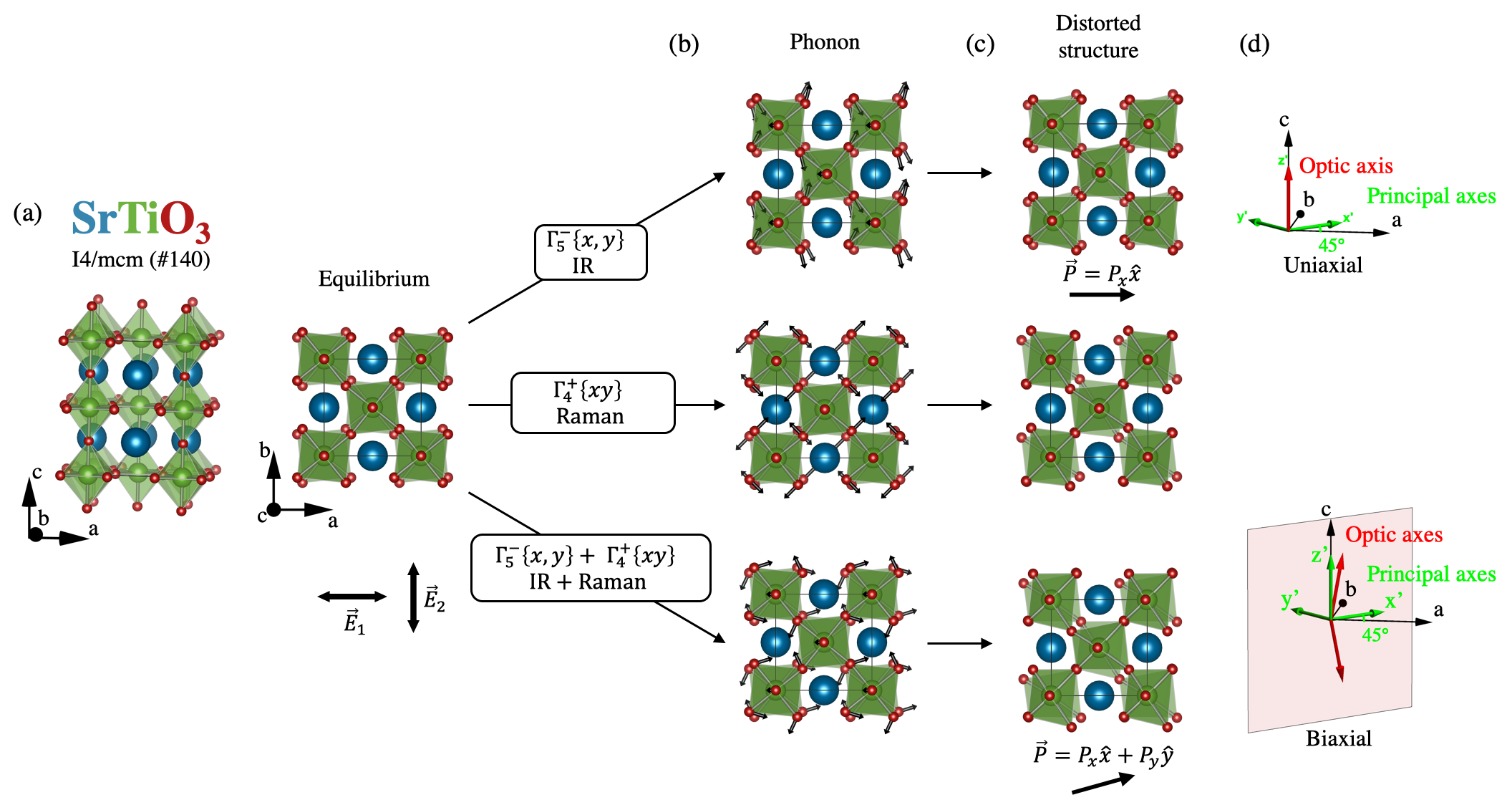}
\caption{Activation of IR and Raman phonons, and symmetry breaking in SrTiO$_3$ through IRRS. (a) Equilibrium crystal structure of SrTiO$_3$ from two perspectives. (b) Schematic representation of the IR, and Raman phonons, and combined IR and Raman phonons. The $\Gamma_5^-$ IR phonon at 16.40 THz, and the $\Gamma_4^+$ Raman phonon at 13.02 THz are shown here. (c) Distorted structure when IR, Raman and the combined IR and Raman phonons are included. The induced polarization is canted by the activation of the $\Gamma_{5}^{-}$ IR phonon polarized in the $x-y$ plane and the $\Gamma_{4}^{+}$ Raman phonon that transforms like $xy$.
(d) Away from equilibrium, the uniaxial nature of SrTiO$_3$ is altered -- splitting and tilting the optical axes. Here, $x',y',z'$ are the principle axes of the distorted crystal with the optical axes canted from $z'$ in the $z'x'$-plane.}
\label{fig:E_IR_Raman_Polarization}
\end{figure*}

The splitting and shift of the optical axes due to the IRRS response in SrTiO$_3$ through excitation of the $\Gamma_{5}^{-}(E_u)$ IR active phonon at 16.40 THz and the $\Gamma_{4}^{+}(B_{2g})$ Raman phonon at 13.02 THz -- both of which primarily involve motion of oxygen atoms -- is shown schematically in \cref{fig:E_IR_Raman_Polarization}. Here the Cartesian coordinates $\{x,y,z\}$ are taken to be parallel with the crystallographic axes $\{a,b,c\}$ and the amplitudes of distortions have been exaggerated for viewing. For light incident along the $z-$axis and polarized in the $x-y$ plane, only $\Gamma_{5}^{-}(E_u)$ IR active phonons can be excited directly. For $\vec{E}_1 \parallel \hat{x}$ the induced polarization $\vec{P}$ through excitation of $\Gamma_5^-(E_u)$ IR phonon is along the $x-$axis (top-panel of \cref{fig:E_IR_Raman_Polarization}(b,c)). Considering only the linear susceptibility, SrTiO$_3$ remains uniaxial (top-panel of \cref{fig:E_IR_Raman_Polarization}(d)). Through IRRS, measured by a second field $\vec{E}_2 \parallel \hat{y}$, motion of the $\Gamma_4^+(B_{2g}) $ is also induced (middle-panel of  \cref{fig:E_IR_Raman_Polarization}(b,c)) but on its own does not induce a dipole and therefore cannot affect the polarization direction. The combined response of these IR and Raman phonons (bottom-panel of  \cref{fig:E_IR_Raman_Polarization}(b,c)) causes the frequency dependent canting of the polarization which is seen in the induced off-diagonal terms in the effective linear susceptibility, as described above. The biaxial optical axes induced by the IRRS are shown in the bottom-panel of  \cref{fig:E_IR_Raman_Polarization}(d), where the off-diagonal components of the effective linear-susceptibility have now forced the direction of the in-plane principle axes (shown as $\{x',y',z'\}$).

We note that the canting of the polarization in SrTiO$_3$ is not \emph{directly} due to the displacement of ions through the motion of the IR and Raman phonons. Rather, the combined motion of the IR and Raman phonons lowers the symmetry of the crystal, this induces changes to the electronic structure of the occupied states, which consequently cants the polarization. Although motion of the ions does not contribute directly to the canting of the polarization in SrTiO$_3$, it is unclear how general this is in arbitrary crystalline systems.

\textbf{Giant refractive index shifts.} When $\vec{E_1}$ is instead resonant with the lowest frequency $\Gamma_5^-(E_u)$
IR active phonon at $2.67\ \text{THz}$,  $\chi^{(3)}$ can be remarkably large -- approaching unity at resonance peaks when $|E_1|$ approaches 1 MV/cm. That is, the shift in the effective linear susceptibility for the infrared resonant Raman response in SrTiO$_3$ can be comparable to the typical linear dielectric response of many optical materials ($\chi^{(1)}\approx 1-10$). \cref{fig:chi3_GM5-_low} shows $\chi^{(3)}_{zxxx}(2;1,-1,2)$ for this case. Here the $\Gamma_5^+(E_g)$ Raman phonons, which have $\{yz,zx\}$ symmetry, are activated in the optical process so the optical changes are again along a perpendicular axis (that is $\vec{E_2}$ is polarized along the $z-$axis). 
The $1.25\ \text{THz}$ and $4.35\ \text{THz}$ $\Gamma_5^+(E_g)$ phonons dominate the IRRS process while the $13.09\ \text{THz}$ $\Gamma_5^+(E_g)$ is not significantly altering the optical response because of its weak coupling to the $2.57\ \text{THz}$ IR phonon.
Difference-frequency (Stokes) and sum-frequency (anti-Stokes) responses are clearly seen in the amplitude and phase of $\chi^{(3)}$ at  ${1.68\ \text{THz}}$, and ${1.42\ \text{THz}}$, and ${7.02\ \text{THz}}$, and ${3.92\ \text{THz}}$, respectively. Interpreting the nonlinear optical changes as intensity dependent susceptibility changes suggests $\delta\chi^{(1)}_{yz}(\omega_2;I_1) \ne 0$ and $\delta\chi^{(1)}_{zx}(\omega_2;I_1) \ne 0$. Focusing on the response above $10\ \text{THz}$ we again note a phase of $\pi/2$ and a fall-off with frequency proportional to $f^{-2}$ showing that the large intensity dependent changes to the real parts of the effective $\chi^{(1)}_{yz}$ and $ \chi^{(1)}_{zx}$ extend far above the region where optical phonons are present.

% % % % % % % % % % % % % % % % % % % % % % % % % % % % % % 
% \Gamma_5^- (x- and y-polarized) IR phonons - Nonlinear coupling
% Clean - Combined
% % % % % % % % % % % % % % % % % % % % % % % % % % % % % %
\begin{table}[h!]
\centering
\begin{tabular}{|l|r|r|r|}
\multicolumn{4}{c}{$\Gamma_5^-(E_u)$ IR phonon at $16.40\ \text{THz}$} \\ 
\hline
Symmetry                & $f(THz)$ & $M(m_u)$ & $b (e/\angstrom)$   \\ \hline \hline

$\Gamma_1^+(A_{1g})$   : $1$            &  4.22    & 16.00    &        0.33         \\ \hline

$\Gamma_2^+(B_{1g})$   : $x^2-y^2$      & 15.21    & 16.00    &       -0.35         \\ \hline

$\Gamma_4^+(B_{2g})$   : $xy$           & 13.02    & 16.02    &       -0.68         \\
                                        &  4.52    & 86.95    &       -0.28         \\ \hline
                       
\multicolumn{4}{c}{$\Gamma_5^-(E_u)$ IR phonon at $2.67\ \text{THz}$} \\ 
\hline
Symmetry                & $f(THz)$ & $M(m_u)$ & $b (e/\angstrom)$   \\ \hline \hline
$\Gamma_5^+(E_{g})$    : $\{yz,zx\}$    & 13.09    & 16.03    &       -0.03         \\
                                        &  4.35    & 80.05    &       -1.33         \\
                                        &  1.25    & 16.25    &        0.26         \\ \hline

\end{tabular}
\caption{Coupling coefficients for SrTiO$_3$ used in \cref{fig:chi3_GM5-_high,fig:chi3_GM5-_low}. The reduced mass and effective charges for the high- and low-frequency $\Gamma_5^-(E_u)$ IR phonons are $M = 16.01\ \mu$ and $\tilde{Z}^*=6.92\ \text{e}$, and $M = 20.12\ \mu$ and $\tilde{Z}^*=14.97\ \text{e}$, respectively.}
\label{table:phonon_table}
\end{table}

\section{Discussion} \label{sec:Discussion}

The findings of Sections \ref{sec:Theoretical_Model} and \ref{sec:Results} suggest a broad relevance to the fields of optics and materials physics.

In optics, IRRS offers a previously unidentified route to strong control of optical material properties. In the analysis of SrTiO$_3$ in Section \ref{sec:Results}, large, broad-frequency changes to the dielectric response were observed in the form of new tensor elements -- in one case reflecting an intensity dependent shift from uniaxial to biaxial symmetry -- as well as the induction of strong absorption and emission peaks, and large refractive index shifts. Extending these ideas to include circularly polarized fields, we might expect IRRS to give strong control of chiral optical response as well. Such large effects were possible via a \textit{triply-resonant} Raman scattering susceptibility through the coupling of IR and Raman phonons mediated by nonlinear contributions to the polarizability, as shown in Eq. (2). 

Whereas it is akin to the conventional electronically mediated resonant Raman response, which couples Raman phonons to virtual electronic dipole transitions through the nonlinear polarizability, IRRS is relevant over a vastly different frequency range relevant to the emergent fields of mid-infrared and THz optics. The theoretical approach we use is quite general and can be used to analyze IRRS in a wide variety of optical materials, including other complex oxides and III-V semiconductors. The main features of a material allowing strong IRRS response are large nonlinear polarizability $b=\frac{\partial \Delta \vec{P}}{\partial Q_{IR} \partial Q_{R}}$ and effective charge $\tilde{Z}$. SrTiO$_3$ has modest values of these parameters, and while the evaluation of other materials is beyond the scope of this article, we might anticipate optical materials with significantly larger $\chi^{(3)}_{b,b}$.

The parameter $b$ can be viewed as the change in the mode-effective charge $\tilde{Z}^*$ due to the Raman phonon (see discussion below Eqn. \ref{eqn:lattice_polarizability}). In SrTiO$_3$ we find that the parameter $b$ is dominated by changes to so-called anomalous charge contributions to the mode-effective charge -- changes to the electronic structure of occupied electronic states. This finding suggests that materials with large anomalous charge should be the focus of the initial search for materials with sizable infrared resonant Raman response. Perovskites are well-known to exhibit large anomalous charges and are therefore a useful theoretical and experimental test ground for IRRS.

The principle of IR resonant modification of $\chi^{(1)}$ via IRRS could be readily extended to explore infrared-driven control of higher-order susceptibilities in both centrosymmetric and non-centrosymmetric crystals where large changes to $\chi^{(2)}_{eff}$, $\chi^{(3)}_{eff}$, \textit{etc.}, may be anticipated due to the dependence of the symmetry and size of these optical constants on Raman phonon displacements. Since some Raman phonons in non-centrosymmetric media have the same symmetry as IR-active phonons, evaluation of IRRS-induced shifts of $\chi^{(2)}$ is considerably more involved. However, such media present a complex response to IR-driven light that might be used to simultaneously tailor linear, electro-optic, and higher order constants through the application of IR-resonant light fields.
For example, in AlN, the $\Gamma_1(A_1)$ optical phonon is simultaneously IR and Raman active, and invariant under all symmetry operations. Extending the discussion of Section \ref{subsec:Raman_Tensor}, \emph{every} optical constant, of every order, will be proportional to the displacement of this phonon while also directly excitable.

We anticipate that the potential utility provided by IRRS through the effects seen in Section \ref{sec:Results}
for optical technologies will rely on three considerations: the applicability of having strong absorption of a control pulse over a short length scale, the possible utility of achieving relative shifts in propagation constants with field polarization, and the degree to which the use of multiple laser frequencies is practical. Through the application of propagating or standing-wave infrared light fields, one might imagine significant applications on the scale of integrated optical devices, \textit{e.g.}, in which side-pumping of a waveguide is used to induce effects such as light-induced wave retardation, polarization ellipse rotation, Bragg reflection, \textit{etc}. 

We note that the optical effects described in this article offer a route to control of optical symmetry akin to that provided by the Pockels effect (linear electro-optic effect), but by a very different physical mechanism. The Pockels effect induces an optical-field dependent shift to the linear electric susceptibility via the action of a strong DC or low-frequency electric field through a nonzero quadratic electric susceptibility, $\delta \chi^{(1)}_{ij} = 2\chi^{(2)}_{ijk}E_k(\omega \approx 0)$. The low-frequency field shifts the electronic distribution from its equilibrium coordination, thus breaking the original symmetry of the electronic potential with respect to new electronic displacements induced by a second electric field. In contrast, IRRS induces a change to the \textit{lattice} from its equilibrium structure, thus also breaking the original symmetry of the electronic potential, as the electronic displacements induced by a second electric field are now with respect to the electronic distribution of the changed lattice. 

Experimental verification of IRRS in mid-IR experiments is most straightforward through observation of an IR-resonant enhancement of Stokes/anti-Stokes Raman peaks. Inspecting \cref{fig:chi3_GM5-_high} it is clear that the spectral features for IRRS and ionic Raman scattering in SrTiO$_3$ are different, allowing for the experimental distinction between the two effects. Additionally, our first-principles evaluation of the coupling terms in SrTiO$_3$ suggest that the resonant enhancement is overwhelmed by IRRS, with many orders of magnitude smaller contribution from ionic Raman scattering, that is, through the anharmonic lattice potential. The degree to which this is a general expectation requires further investigation and broad sampling of crystalline optical materials that is beyond the scope of this study. Additionally, the frequency dependent fall-off at high-frequency is different between the two pathways with IRRS falling-off as $f^{-2}$ and ionic Raman scattering falling of as $f^{-4}$, providing another measurable confirmation of IRRS. We note that additional cross-coupling Raman pathways exist and have provided expressions for the polarization and susceptibility in the most general case applied to centrosymmetric crystals in
\cref{app_eqn:dipole_complete_solution,app_eqn:dipole_solution,app_eqn:third-order_susceptibility_general,app_eqn:third-order_susceptibility_two_fields}.
With these findings our view is that, while the anharmonic lattice potential is certainly contributing to the material response, the nonlinear lattice polarizability is providing a primary mechanism for change of the dielectric response and therefore experimental interaction with, and control of, the material’s optical properties.

\begin{figure}[h!]
\centering
\includegraphics[scale=0.32]{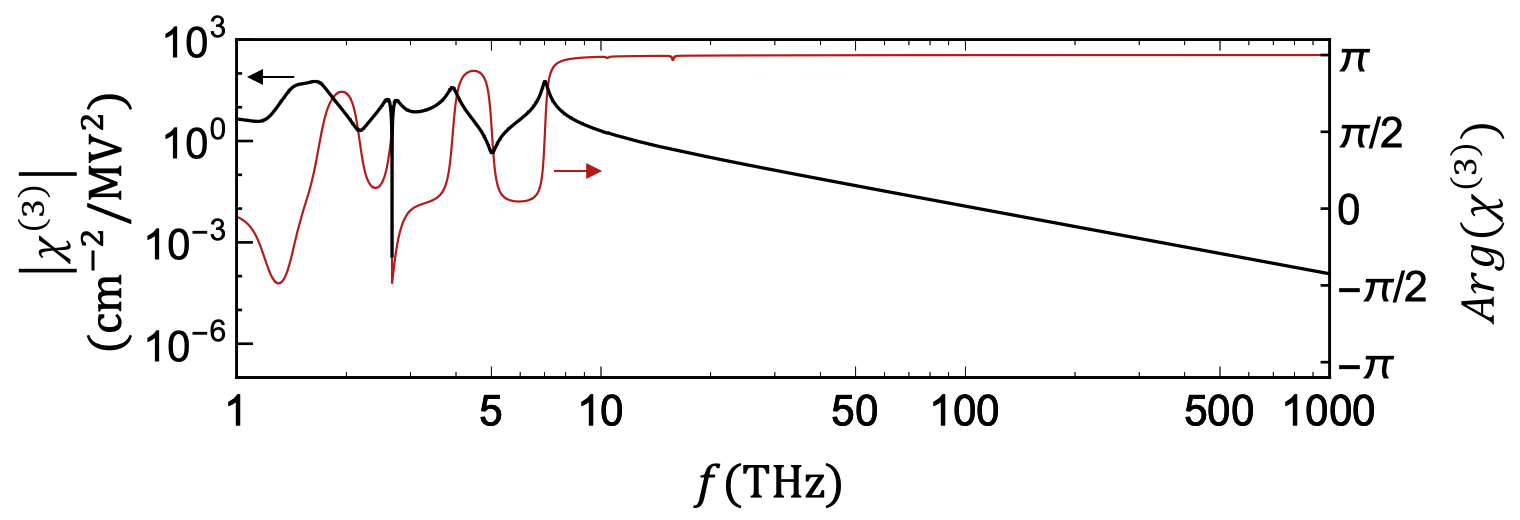}
\caption{Simulated magnitude (black) and phase (red) of the third-order susceptibility due to IRRS, $\chi^{(3)}_{b,b}$, in SrTiO$_3$ when the lowest frequency $\Gamma_5^-(E_u)$ IR phonon is resonantly excited (${f_{IR}=2.67\ \text{THz}}$). $\chi^{(3)}_{zyxx}(2;1,-1,2)$ is now facilitated by the $\Gamma_5^+(E_g)$ Raman phonons. The difference- and sum-frequencies are now ${1.68\ \text{THz}}$, and ${1.42\ \text{THz}}$, and ${7.02\ \text{THz}}$, and ${3.92\ \text{THz}}$, respectively. The coupling to the ${13.09\ \text{THz}}$ Raman phonon is weak and does not alter the response significantly.}
\label{fig:chi3_GM5-_low}
\end{figure}

The IRRS process between the IR-active $\Gamma_5^-$ phonons and the Raman-active $\Gamma_5^+$ phonons (\cref{fig:chi3_GM5-_low}) provides a symmetry constrained strategy for comparison of IRRS and ionic Raman scattering. The induced polarization along the $z-$axis measured by $E_2$ in IRRS through the nonlinear lattice polarizability $\Delta P \propto Q_R Q_{IR}$ is not allowed in conventional ionic Raman scattering. In ionic Raman scattering, an additional IR-active phonon polarized along the $z-$axis ($\Gamma_3^-$) is required to mediate the interaction through an anharmonic lattice potential term $\propto Q_{R}Q_{IR1}Q_{IR2}$. This leads to a distinction in the resonance conditions for ionic Raman scattering and IRRS, with $f_{IR} \pm f_2 = f_R$ in the IRRS case and $f_R = f_{IR1} \pm f_{IR2}$ in the ionic Raman scattering case, where $f_{IR1,2}$ describe the $\Gamma_5^-$ and $\Gamma_3^-$ phonons (see the right column of \cref{fig:comparison}). The peak at 7.02 THz in \cref{fig:chi3_GM5-_low} should experimentally distinguish the two effects since there is no IR resonance in SrTiO$_3$ polarized along the $z-$axis at that frequency.

For materials science, IRRS also provides a new paradigm for light-induced structural control and analysis. Notably, it provides a flexible strategy for structural control in addition to the nonlinear phononics effect mediated by the anharmonic potential and therefore expands the toolbox for ultrafast structural control of novel functional materials. The unidirectional force induced by the anharmonic potential in response to intense resonant IR excitation is also found through IRRS. This can be seen from the first line of \cref{eqn:chi_3_nonlinear_ionic_dipole} where the response of the Raman phonon is frequency independent. In \cref{app_eqn:first_order_coordinates} perturbative results for the Raman coordinate show that the resonant conditions for unidirectional Raman displacement are different between the two effects allowing for experimental separation of the two mechanisms. 
In addition to the differing resonant conditions between IRRS and ionic Raman scattering, the appearance of $b$ and $B$ in the equations of motion (\cref{eqn:equations_of_motion_all}) suggest the two effects will have different timescales in the transient response, with IRRS depending sensitively on the timescale of the pulse, and ionic Raman scattering potentially depending sensitively on the lifetime of the excited IR phonon. \newtext{For structural verification of the IRRS pathway, we predict that future mid-IR pump/X-ray-probe experiments will find a unidirectional displacement of Raman phonons that initially overshoots/undershoots the long-lived quasi-static displacement for a time-scale comparable to the excitatory mid-IR pump pulse.} We note, additionally, that the strong dependence of optical constants on light-driven phonon displacements via IRRS suggest another possible physical mechanism for the spectral features attributed to the nonlinear phononics effect in optical probe measurements.

While the anharmonic potential pathway to displacements of Raman phonons is restricted to frequencies defined by the IR phonon resonance frequencies through the effective force $B Q_{IR}^2$, the force induced by the IRRS pathway, $b Q_{IR} E$, depends continuously on the electric field of a second laser. This provides a flexible, polarization dependent pathway for lattice excitation. We anticipate that in addition to the optical symmetry considerations and breaking described in Sections \ref{sec:Symmetry_Considerations} and \ref{sec:Results}, IRRS and its extension to noncentrosymmetric crystals, through the use of multiple lasers, will provide tailored transient out-of-equilibrium structural symmetry control.

% % % % % % % % % % % % % % % % % % % % % % % % % % % % % % 
% Acknowledgments
% % % % % % % % % % % % % % % % % % % % % % % % % % % % % % 
\begin{acknowledgments}
This work was supported by the Cornell Center for Materials Research with funding from the NSF MRSEC program (DMR-1719875). Computational resources were provided by the Cornell Center for Advanced Computing.
\end{acknowledgments}

% % % % % % % % % % % % % % % % % % % % % % % % % % % % % % 
% Appendices
% % % % % % % % % % % % % % % % % % % % % % % % % % % % % % 
\appendix

\section{Perturbative results for the phonon coordinates, polarization, and susceptibility}

In this Appendix we derive perturbative results for the electronic and phonon coordinates, as well as expressions for the polarization and third-order susceptibility including all possible coupling pathways good to second-order in the coupling parameters $b$, $\beta$, and $B$.

We define the perturbation expansion in the parameters $b$, $\beta$, and $B$ for each coordinate $Q_{\sigma}$ as
\begin{equation}
    \label{app_eqn:cross-coupling_pert_expansion}
    \begin{split}
        Q_{\sigma}  &= Q_{\sigma}^{(000)} \\
                    &+ \beta Q_{\sigma}^{(100)} + b Q_{\sigma}^{(010)} + B Q_{\sigma}^{(001)} \\
                    &+ \beta b Q_{\sigma}^{(110)} + \beta B  Q_{\sigma}^{(101)} + b B  Q_{\sigma}^{(011)} \\
                    &+ \beta^2 Q_{\sigma}^{(200)} + b^2 Q_{\sigma}^{(020)} + B^2 Q_{\sigma}^{(002)} + ...\\
    \end{split}
\end{equation}
\noindent
where $\sigma = \{e,IR,R\}$. 

To derive the perturbation theory results we define electric field as a general multi-component field of the form
\begin{equation}
    \label{app_eqn:electric_field_expansion}
    E(t)=\frac{1}{2}\sum_{n=\pm 1}^{\pm N}E_n e^{-i\omega_n t}
\end{equation}
\noindent
where we take $E_{-n} = E_n^*$ and $\omega_n = - \omega_{-n}$ to enforce real fields. Since the zeroth-order solution is proportional to the electric field, we expect a similar expansion coefficients, but the first- and second-order solutions are driven by higher-order expressions in the electric field. To accomodate this we expand $Q_{\sigma}$ as

\begin{align}
    Q^{(lmn)}_{\sigma} &= \frac{1}{2}\sum_{n} e^{-i \omega_{n} t} Q_{\sigma,n}, & l+m+n=0 \nonumber \\
    Q^{(lmn)}_{\sigma} &= \frac{1}{4}\sum_{mn} e^{-i \omega_{mn} t} Q_{\sigma,mn}, & l+m+n=1 \label{app_eqn:temporal_dependence} \\ 
    Q^{(lmn)}_{\sigma} &= \frac{1}{8}\sum_{lmn} e^{-i \omega_{lmn} t} Q_{\sigma,lmn}, & l+m+n=2 \nonumber
\end{align}
\noindent
where the frequencies are constrained by $\omega_{mn} = \omega_m + \omega_n = - \omega_{-m-n}$ and $\omega_{lmn} = \omega_l + \omega_m + \omega_n = - \omega_{-l-m-n}$, and the expansion coefficients $Q_{\sigma,-m-n} = Q_{\sigma,mn}^*$, and $Q_{\sigma,-l-m-n} = Q_{\sigma,lmn}^*$ are defined to enforce real response functions.

After plugging the expansion from \cref{app_eqn:cross-coupling_pert_expansion,app_eqn:electric_field_expansion,app_eqn:temporal_dependence} into \cref{eqn:equations_of_motion_all} and collecting terms we iteratively find the perturbation coefficients. The zeroth-order results are
\begin{equation}
    \label{app_eqn:zeroth_order_coordinates}
    \begin{split}
        Q_{e,n}^{(000)}  &= \zeta   G^{e}_{n} E_n \\
        Q_{IR,n}^{(000)} &= \tilde{Z}^* G^{IR}_{n} E_n .\\
    \end{split}
\end{equation}
\noindent
From which we find the following first-order expressions
\begin{equation}
    \label{app_eqn:first_order_coordinates}
    \begin{split}
        Q_{R,mn}^{(100)}   &= \zeta G^{R}_{mn} G^{e}_{n} E_m E_n \\
        Q_{R,mn}^{(010)}   &= \tilde{Z}^{*} G^{R}_{mn} G^{IR}_n E_m E_n \\
        Q_{R,mn}^{(001)}   &= \tilde{Z}^{*2} G^{R}_{mn} G^{IR}_m G^{IR}_n E_m E_n
    \end{split}
\end{equation}
\noindent
and the second-order expression

\newpage
\begin{widetext}
\begin{equation}
    \label{app_eqn:second_order_coordinates}
    \begin{split}
        Q_{IR,lmn}^{(011)} &= \tilde{Z}^{*2} G^{IR}_{lmn} \left(G^{R}_{mn} G^{IR}_{m} G^{IR}_{n} + 2 G^{IR}_{l}  G^{R}_{mn} G^{IR}_{n} \right) E_l E_m E_n \\
        Q_{e,lmn}^{(200)}  &= \zeta G^{e}_{lmn} G^{R}_{mn} G^{e}_{n} E_l E_m E_n \\
        Q_{IR,lmn}^{(020)} &= \tilde{Z}^{*} G^{IR}_{lmn} G^{R}_{mn} G^{IR}_{n} E_l E_m E_n \\
        Q_{IR,lmn}^{(002)} &= 2 \tilde{Z}^{*3} G^{IR}_{lmn} G^{IR}_{l} G^{R}_{mn} G^{IR}_m G^{IR}_n E_l E_m E_n.
    \end{split}
\end{equation}
\end{widetext}

\noindent
These results can be inserted into \cref{eqn:lattice_polarizability,eqn:electronic_polarizability} to find the polarization to second-order in the perturbed quantities,
\begin{widetext}
\begin{equation} 
    \label{app_eqn:dipole_complete_solution}
    \begin{split}
        \epsilon_0 V \Delta P 
                 &= \zeta \left( Q_{e}^{(000)} + \beta b Q_{e}^{(110)} + \beta B Q_{e}^{(101)} + \beta^2 Q_{e}^{(200)} \right) \\
                 &+ \beta \left( \beta Q^{(100)}_R + b Q^{(010)}_R + B Q^{(001)}_R \right) Q_{e}^{(000)} \\
                 &+ \tilde{Z}^{*} \left( Q^{(000)}_{IR} + \beta b Q^{(110)}_{IR} + b B Q^{(011)}_{IR} + b^2 Q^{(020)}_{IR} + B^2 Q^{(002)}_{IR} \right) \\
                 &+ b \left( \beta Q^{(100)}_R + b Q^{(010)}_R + B Q^{(001)}_R \right) \\
                 \\
                 &= \zeta^2 \frac{1}{2} \sum_n G^{e}_{n} e^{-i\omega_n t} E_n \\ 
                 &+ \tilde{Z}^{*2} \frac{1}{2} \sum_n G^{IR}_{n} e^{-i\omega_n t} E_n \\
                 &+ \zeta^2 \beta^2 \frac{1}{8} \sum_{lmn} \left( G^{e}_{l} G^{R}_{mn} G^{e}_{n} + G^{e}_{lmn} G^{R}_{mn} G^{e}_{n} \right) e^{-i\omega_{lmn} t} E_l E_m E_n \\
                 &+ \tilde{Z}^{*2} b^2 \frac{1}{8} \sum_{lmn} \left( G^{IR}_{l} G^{R}_{mn} G^{IR}_{n} + G^{IR}_{lmn} G^{R}_{mn} G^{IR}_{n} \right) e^{-i\omega_{lmn} t} E_l E_m E_n \\
                 &+ \tilde{Z}^{*4} B^2 \frac{1}{8} \sum_{lmn} \left( 2 G^{IR}_{lmn} G^{IR}_{l} G^{R}_{mn} G^{IR}_m G^{IR}_n \right) e^{-i\omega_{lmn} t} E_l E_m E_n \\
                 &+ \zeta \beta \tilde{Z}^{*} b \frac{1}{8} \sum_{lmn} \left( G^{e}_{l} G^{R}_{mn} G^{IR}_{n} + G^{e}_{lmn} G^{R}_{mn} G^{IR}_{n} + G^{IR}_{l} G^{R}_{mn} G^{e}_{n} + G^{IR}_{lmn} G^{R}_{mn} G^{e}_{n} \right) e^{-i\omega_{lmn} t} E_l E_m E_n \\
                 &+ \zeta \beta \tilde{Z}^{*2} B \frac{1}{8} \sum_{lmn} \left( G^{e}_{lmn} G^{R}_{mn} G^{IR}_{m} G^{IR}_{n} + G^{e}_{l} G^{R}_{mn} G^{IR}_{m} G^{IR}_{n} \right) e^{-i\omega_{lmn} t} E_l E_m E_n \\
                 &+ \tilde{Z}^{*3} b B \frac{1}{8} \sum_{lmn} \left( G^{IR}_{l} G^{R}_{mn} G^{IR}_{m} G^{IR}_{n} + G^{IR}_{lmn} G^{R}_{mn} G^{IR}_{m} G^{IR}_{n} + 2 G^{IR}_{lmn} G^{IR}_{l} G^{R}_{mn} G^{IR}_{n} \right) e^{-i\omega_{lmn} t} E_l E_m E_n \\
    \end{split}
\end{equation}
\end{widetext}
\noindent
where $V$ is the unit cell volume and $\epsilon_0$ is the free-space permitivity. From this result we can construct the linear and cubic susceptibilites but we focus on the case of a wide bandgap insulator where the electronic resonance $\omega_e$ is much larger than all working frequencies. In this case we can make the approximation that {$G^{e}_{n} = G^{e}_{lmn} = G^e(\omega \rightarrow 0)$} and redefine {$\xi_R = \sqrt{2} \zeta \beta G^e(\omega = 0)$} and {$\chi^{(1)}_{e,0} = \zeta^2 G^{e}(\omega=0)/\epsilon_0 V$}. We are then left with the polarization of
\begin{widetext}
\begin{equation} 
    \label{app_eqn:dipole_solution}
    \begin{split}
        \epsilon_0 V \Delta P 
                 &= \frac{1}{2} \epsilon_0 V \chi^{(1)}_{e,0} \sum_n e^{-i\omega_n t} E_n 
                 + \frac{1}{2} \tilde{Z}^{*2} \sum_n G^{IR}_{n} e^{-i\omega_n t} E_n \\
                 &+ \frac{1}{8} \xi_R^2 \sum_{lmn} G^{R}_{mn} e^{-i\omega_{lmn} t} E_l E_m E_n \\
                 &+ \frac{1}{8} \tilde{Z}^{*2} b^2 \sum_{lmn} \left( G^{IR}_{l} G^{R}_{mn} G^{IR}_{n} + G^{IR}_{lmn} G^{R}_{mn} G^{IR}_{n} \right) e^{-i\omega_{lmn} t} E_l E_m E_n \\
                 &+ \frac{1}{8} \tilde{Z}^{*4} B^2 \sum_{lmn} \left( 2 G^{IR}_{lmn} G^{IR}_{l} G^{R}_{mn} G^{IR}_m G^{IR}_n \right) e^{-i\omega_{lmn} t} E_l E_m E_n \\
                 &+ \frac{1}{8} \frac{\xi_R \tilde{Z}^{*} b}{\sqrt{2}} \sum_{lmn} \left( 2 G^{R}_{mn} G^{IR}_{n} + G^{IR}_{l} G^{R}_{mn} + G^{IR}_{lmn} G^{R}_{mn} \right) e^{-i\omega_{lmn} t} E_l E_m E_n \\
                 &+ \frac{1}{8} \frac{\xi_R \tilde{Z}^{*2} B}{\sqrt{2}} \sum_{lmn} \left( 2 G^{R}_{mn} G^{IR}_{m} G^{IR}_{n} \right) e^{-i\omega_{lmn} t} E_l E_m E_n \\
                 &+ \frac{1}{8} \tilde{Z}^{*3} b B \sum_{lmn} \left( G^{IR}_{l} G^{R}_{mn} G^{IR}_{m} G^{IR}_{n} + G^{IR}_{lmn} G^{R}_{mn} G^{IR}_{m} G^{IR}_{n} + 2 G^{IR}_{lmn} G^{IR}_{l} G^{R}_{mn} G^{IR}_{n} \right) e^{-i\omega_{lmn} t} E_l E_m E_n \\
    \end{split}
\end{equation}
\end{widetext}

Collecting terms proportional to $E_l E_m E_n$ gives the $(lmn)^{th}$ components of the third-order susceptibilities. We find

\begin{widetext}
\begin{equation}
    \label{app_eqn:third-order_susceptibility_general}
    \begin{split}
        \chi^{(3)}_{\xi_R,\xi_R,lmn}      &= \xi_R^2 G^{R}_{mn} \\
        \chi^{(3)}_{b,b,lmn}     &= \tilde{Z}^{*2} b^2 \left( G^{IR}_{l} G^{R}_{mn} G^{IR}_{n} + G^{IR}_{lmn} G^{R}_{mn} G^{IR}_{n} \right) \\
        \chi^{(3)}_{B,B,lmn}    &= \tilde{Z}^{*4} B^2 \left( 2 G^{IR}_{lmn} G^{IR}_{l} G^{R}_{mn} G^{IR}_m G^{IR}_n \right) \\
        \chi^{(3)}_{\xi_R,b,lmn}   &= \frac{\xi_R \tilde{Z}^{*} b}{\sqrt{2}} \left( 2 G^{R}_{mn} G^{IR}_{n} + G^{IR}_{l} G^{R}_{mn} + G^{IR}_{lmn} G^{R}_{mn} \right) \\
        \chi^{(3)}_{\xi_R,B,lmn}  &= \frac{\xi_R \tilde{Z}^{*2} B}{\sqrt{2}} \left(2 G^{R}_{mn} G^{IR}_{m} G^{IR}_{n} \right) \\
        \chi^{(3)}_{b,B,lmn} &= \tilde{Z}^{*3} b B \left( G^{IR}_{l} G^{R}_{mn} G^{IR}_{m} G^{IR}_{n} + G^{IR}_{lmn} G^{R}_{mn} G^{IR}_{m} G^{IR}_{n} + 2 G^{IR}_{lmn} G^{IR}_{l} G^{R}_{mn} G^{IR}_{n} \right)
        \end{split}
\end{equation}
\end{widetext}
\noindent
In these expressions we have explicitly separated the possible contributions to the third-order susceptibilites. The first three terms describe the conventional electronic response, the IRRS contribution from $b$, and the response from $B$, respectively. The last three terms describe cross-coupling between the $b$, $\beta$, and $B$ pathways.

The expressions derived above can be used for an arbitrary electric field. In order to study the IRRS for two frequencies of light, we must account for the all combinatoric possibilities in the sums over $l$, $m$, and $n$. When the electric field is defined as it is in the main text
\begin{equation*}
    E (t)= \frac{1}{2} \left( E_1 e^{-i \omega_1 t} + E_{-1} e^{i \omega_1 t} + E_2 e^{-i \omega_2 t} +  E_{-2} e^{i \omega_2 t} \right)
\end{equation*}
\noindent
we find the following expressions for the cubic susceptibilities contributing to {$\Delta P = \chi^{(3)}(2;1,-1,2) |E_1|^2 E_2 e^{-i\omega_1 t}$}
\begin{widetext}
\begin{equation}
    \label{app_eqn:third-order_susceptibility_two_fields}
    \begin{split}
        \chi^{(3)}_{\xi_R,\xi_R}(2;1,-1,2) 
            = \frac{\xi_R^2}{\epsilon_0 V} \times & \left\{4 G^R(\omega_{mn}=0) \right. \\
            &+      2 G^{R}(-\omega_1 + \omega_2) \\
            &+\left. 2 G^{R}( \omega_1 + \omega_2) \right\} \\
        \chi^{(3)}_{\xi_R,b}(2;1,-1,2) 
            = \frac{\xi_R \tilde{Z}^{*} b}{\sqrt{2}\epsilon_0 V} \times & \left\{ \left[2 \left(G^{IR}_1 + G^{IR}_{-1} + G^{IR}_2 + G^{IR}_{-2} \right) + 8 G^{IR}_2\right]G^R(\omega_{mn}=0) \right. \\
            &+ \left[ 4G^{IR}_2 + 2\left( G^{IR}_1 + G^{IR}_{-1} \right) \right] G^{R}(-\omega_1 + \omega_2) \\
            &\left.+ \left[ 4G^{IR}_2 + 2\left( G^{IR}_1 + G^{IR}_{-1} \right) \right] G^{R}( \omega_1 + \omega_2)  \right\} \\
        \chi^{(3)}_{b,b}(2;1,-1,2) 
            = \frac{(b \tilde{Z}^{*2})}{\epsilon_0 V} \times & \left\{ 2 \left(G^{IR}_1 + G^{IR}_{-1} + G^{IR}_2 + G^{IR}_{-2} \right) G^{IR}_2 G^R(\omega_{mn}=0) \right. \\
            &+ \left( G^{IR}_2 G^{IR}_2 + G^{IR}_1 G^{IR}_2 + G^{IR}_2 G^{IR}_{-1} + G^{IR}_1  G^{IR}_{-1} \right) G^{R}(-\omega_1 + \omega_2) \\
            &\left.+ \left( G^{IR}_2 G^{IR}_2 + G^{IR}_1 G^{IR}_2 + G^{IR}_2 G^{IR}_{-1} + G^{IR}_1  G^{IR}_{-1} \right) G^{R}(\omega_1 + \omega_2)   \right\}\\
        \chi^{(3)}_{\xi_R,B}(2;1,-1,2) 
            = \frac{\xi_R \tilde{Z}^{*2} B}{\sqrt{2}\epsilon_0 V} \times & \left\{ 4 \left(G^{IR}_1 G^{IR}_{-1} + G^{IR}_2 G^{IR}_{-2} \right) G^R(\omega_{mn}=0) \right. \\
            &+ 4 \left( G^{IR}_{-1} G^{IR}_2 \right) G^{R}(-\omega_1 + \omega_2) \\
            &\left.+ 4 \left( G^{IR}_{1} G^{IR}_2 \right) G^{R}(\omega_1 + \omega_2)  \right\} \\
        \chi^{(3)}_{b,B}(2;1,-1,2) 
            = \frac{\tilde{Z}^{*3} b B}{\epsilon_0 V} \times & \left\{ 2 \left(G^{IR}_1 G^{IR}_{-1} + G^{IR}_{1} G^{IR}_{2} +  G^{IR}_{-1} G^{IR}_{2} + G^{IR}_2 G^{IR}_{2} + 2 G^{IR}_{2} G^{IR}_{-2} \right)G^{IR}_2 G^R(\omega_{mn}=0) \right. \\
            &+ 2 \left( 2 G^{IR}_{1} G^{IR}_{-1} + G^{IR}_{1} G^{IR}_{2} + G^{IR}_{-1} G^{IR}_{2} \right) G^{IR}_{2} G^{R}(-\omega_1 + \omega_2) \\
            &\left.+ 2 \left( 2 G^{IR}_{1} G^{IR}_{-1} + G^{IR}_{1} G^{IR}_{2} + G^{IR}_{-1} G^{IR}_{2} \right) G^{IR}_{2} G^{R}(\omega_1 + \omega_2) \right\}\\
        \chi^{(3)}_{B,B}(2;1,-1,2) 
            = \frac{\tilde{Z}^{*4} B^2}{\epsilon_0 V} \times & \left\{ 2 \left(G^{IR}_1 G^{IR}_{-1} + G^{IR}_{2} G^{IR}_{-2}\right)(G^{IR}_2)^2 G^R(\omega_{mn}=0) \right. \\
            &+ 2 \left(G^{IR}_1 G^{IR}_{-1}\right)(G^{IR}_2)^2 G^R(-\omega_1+\omega_2)  \\
            &\left.+ 2 \left(G^{IR}_1 G^{IR}_{-1}\right)(G^{IR}_2)^2 G^R(\omega_1+\omega_2) \right\}  \\
    \end{split}
\end{equation}
\end{widetext}

% % % % % % % % % % % % % % % % % % % % % % % % % % % % % % 
% Bibliography
% % % % % % % % % % % % % % % % % % % % % % % % % % % % % % 
\bibliography{bibliography}

\pagebreak
% % % % % % % % % % % % % % % % % % % % % % % % % % % % % % 
% Figures
% % % % % % % % % % % % % % % % % % % % % % % % % % % % % % 

% % % % % % % % % % % % % % % % % % % % % % % % % % % % % % % % % % % 
% Begin Figure 1 - nonlinear dipole moments
% % % % % % % % % % % % % % % % % % % % % % % % % % % % % % % % % % % 
% % % % % % % % % % % % % % % % % % % % % % % % % % % % % % % % % % % 
% End Figure 1 - nonlinear dipole moments
% % % % % % % % % % % % % % % % % % % % % % % % % % % % % % % % % % % 

% % % % % % % % % % % % % % % % % % % % % % % % % % % % % % % % % % % 
% Begin Figure 2 - Driving terms and nonlinear polarizability
% % % % % % % % % % % % % % % % % % % % % % % % % % % % % % % % % % %
% % % % % % % % % % % % % % % % % % % % % % % % % % % % % % % % % % % 
% End Figure 2 - Driving terms and nonlinear polarizability
% % % % % % % % % % % % % % % % % % % % % % % % % % % % % % % % % % %

% % % % % % % % % % % % % % % % % % % % % % % % % % % % % % % % % % % 
% Begin Figure 3 - chi^{(3)} schematic
% % % % % % % % % % % % % % % % % % % % % % % % % % % % % % % % % % %
% % % % % % % % % % % % % % % % % % % % % % % % % % % % % % % % % % % 
% End Figure 3 - chi^{(3)} schematic
% % % % % % % % % % % % % % % % % % % % % % % % % % % % % % % % % % %

% % % % % % % % % % % % % % % % % % % % % % % % % % % % % % % % % % % 
% Begin Figure 4 - chi^{(3)} high frequency GM5-
% % % % % % % % % % % % % % % % % % % % % % % % % % % % % % % % % % %
% % % % % % % % % % % % % % % % % % % % % % % % % % % % % % % % % % % 
% End Figure 4 - chi^{(3)} high frequency GM5-
% % % % % % % % % % % % % % % % % % % % % % % % % % % % % % % % % % %

% % % % % % % % % % % % % % % % % % % % % % % % % % % % % % % % % % % 
% Begin Figure 5 - Efield, IR, Raman, and polarization in SrTiO3
% % % % % % % % % % % % % % % % % % % % % % % % % % % % % % % % % % %
% % % % % % % % % % % % % % % % % % % % % % % % % % % % % % % % % % % 
% End Figure 5 - Efield, IR, Raman, and polarization in SrTiO3
% % % % % % % % % % % % % % % % % % % % % % % % % % % % % % % % % % %

% % % % % % % % % % % % % % % % % % % % % % % % % % % % % % % % % % % 
% Begin Figure 6 - chi^{(3)} low frequency GM5-
% % % % % % % % % % % % % % % % % % % % % % % % % % % % % % % % % % %
% % % % % % % % % % % % % % % % % % % % % % % % % % % % % % % % % % % 
% End Figure 6 - chi^{(3)} low frequency GM5-
% % % % % % % % % % % % % % % % % % % % % % % % % % % % % % % % % % %

\end{document}